\documentclass[epj]{svjour}
\usepackage{graphics}
\def\be{\begin{equation}} \def\ee{\end{equation}} \def\bea{\begin{eqnarray}}
\def\eea{\end{eqnarray}} \def\nnb{\nonumber}
\begin{document}
\title{
Cluster effective field theory and nuclear reactions
}
\author{
Shung-Ichi Ando 
}                     
%
%
\institute{
School of Mechanical and ICT convergence engineering,
Sunmoon University,
Asan, Chungnam 31460,
Republic of Korea
}
\date{Version on November 3, 2020} 
%
\abstract{
An effective field theory (EFT) for a nuclear reaction 
at low energies is studied.
The astrophysical $S$-factor of radiative $\alpha$ capture 
on $^{12}$C at the Gamow-peak energy,
$T_G=0.3$~MeV, is a fundamental quantity in nuclear-astrophysics,
and we construct an EFT for the reaction.
To fix parameters appearing in the effective Lagrangian,
the EFT is applied to the study for three reactions:
elastic $\alpha$-$^{12}$C scattering at low energies, 
$E1$ transition of radiative $\alpha$ capture on $^{12}$C,
and $\beta$ delayed $\alpha$ emission from $^{16}$N. 
We report an estimate of the $S_{E1}$-factor of the reaction
through the $E1$ transition at $T_G$ by employing the EFT.
We also discuss applications of EFTs to nuclear reactions at low energies.
} 
\maketitle
\section{Introduction}
\label{intro}

In the late 1970s, a phenomenological Lagrangian method is suggested
by Weinberg~\cite{w-pa79} as an alternative of current algebra
to calculate a hadronic matrix element at low energies. 
Skillful techniques are required 
for calculations of the current algebra 
to derive a hadronic reaction amplitude, e.g., 
the electro pion production on a nucleon, 
considering CVC and PCAC and satisfying the results
from the low energy theorem and the relations among vertex functions
required by Ward-Takahashi identities~\cite{o-prc92,o-prc93}. 
Those results, including the radiative and non-radiative 
vector and axial-vector matrix elements of nucleon, nonetheless, are
straightforwardly calculated from the chiral Lagrangian,
which embodies the chiral $SU(2)\times SU(2)$ symmetry 
and the explicit symmetry breaking patterns,
in heavy-baryon chiral perturbation 
theory~\cite{bklm-pr94,am-plb98,amk-prc01}.
The idea of the method has been developed in various ways, 
and the framework of those methods are now known as 
effective field theories (EFTs). 
(For a review of EFTs, see, e.g., 
Refs.~\cite{bv-arnps02,bh-pr06,m-15,hjp-17,hkvk-19}.)
In the present work, we discuss a construction of an EFT 
for nuclear reactions at low energies 
based on our recent works~\cite{a-epja16,a-prc18,a-jkps18,ya-19,a-18}. 

When one studies a reaction by means of EFT,
one may expect the characteristics for an EFT listed below:
(1) It is a model independent approach. 
(2) One needs to introduce a momentum scale 
to separate relevant degrees of freedom
at low energies from irrelevant degrees of freedom at high energies. 
(3) The theory provides us a perturbative expansion scheme around a 
specified theoretical limit; 
counting rules in powers of $Q/\Lambda_H$, 
where $Q$ denotes a typical momentum scale of a reaction and $\Lambda_H$
does a high momentum scale,
will be available. By using the counting rules,
one can expand the amplitude order by order, and, 
up to a given order, one has a finite number of diagrams to calculate. 
(4) Coefficients appearing in an effective Lagrangian 
may not be constrained by its mother theory but can be fixed by using
experimental data. 

One might regard the construction of an EFT for a nuclear reaction 
as a formidable task
because a nucleus is a many-body nucleon system
and its structure is not still easily described from first principles.
At low energies, e.g., a long wavelength limit for an external probe,
however, 
an amplitude constructed from an effective Lagrangian, which is 
represented in terms of relevant low energy degrees of freedom 
and embodies symmetry requirements, exhibits a specific expression
in the low energy limit and may describe well the reaction through
the external probe. 
This might be a consequence 
of the low energy theorem~\cite{em-cnpp95} for nuclear reactions.
As discussed above, 
to construct an EFT for a nuclear reaction is possible; 
the important two conditions are (2) a clear separation scale 
and (4) an availability of experimental data for a reaction. 
For application in nuclear-astrophysics, nonetheless,
one encounters an additional task.
Because experimental data are available at relatively high energies
due to the Coulomb barrier, 
it is necessary to work out an extrapolation of a reaction rate to 
a low energy.

In the following sections, we discuss and evaluate the $S_{E1}$ factor
of radiative $\alpha$ capture on $^{12}$C at stellar energy 
through the $E1$ transition by constructing an EFT~\cite{a-18}.
Our approach is rather new compared to popular methods
for nuclear-astrophysics studies, such as $R$-matrix 
or $K$-matrix approach~\cite{lt-rmp58} and potential models. 
Many studies of the $E1$ transition of radiative $\alpha$ capture on 
$^{12}$C have actually been carried out by using the conventional methods, 
in which a crucial observation has been made:
The importance of indirect measurement of a reaction, 
$\beta$ delayed $\alpha$ emission from $^{16}$N, 
was pointed out~\cite{bb-npa06,detal-17}. 
An interference between a subthreshold state 
and a resonant state (also including so called background levels) 
for $l=1$ channel is sensitive to a secondary peak 
at lower energy side of the $\alpha$ energy
spectrum from $\beta$-delayed $\alpha$ emission from $^{16}$N. 
That is an essential input to deduce the $S_{E1}$ factor when one employs
the $R$-matrix approach. 
Meanwhile, this feature appears quite differently in our approach: 
the subthreshold and resonant states for the $l=1$ channel are represented
within a single dressed $^{16}$O propagator while the secondary peak  
of the $\alpha$ energy spectrum from $\beta$ delayed $\alpha$ emission 
from $^{16}$N is described by an interference between the amplitudes 
from a pole diagram and a non-pole one.
Because one might argue about the applicability of the present approach to 
studies of nuclear reactions (e.g., because of no interference between
the subthreshold and resonant states for the $l=1$ channel), 
the related reactions to estimate the $S_{E1}$-factor should 
be studied simultaneously using the same formalism. 
In addition, because the $S_{E1}$-factor has recently been estimated 
by using mainly a single method, the $R$-matrix approach, 
the model dependence of the method is worth questioning. 
An examination employed by another method would be called for. 

In the present work, we review a series of the calculations for the $E1$ 
transition of radiative $\alpha$ capture on $^{12}$C by constructing
an EFT, in which
three reactions, elastic $\alpha$-$^{12}$C 
scattering~\cite{a-epja16,a-prc18,a-jkps18,ya-19},
$S_{E1}$ factor of the radiative $\alpha$ capture on $^{12}$C~\cite{a-18}, 
and $\beta$ delayed $\alpha$ emission from $^{16}$N, are discussed. 
The construction of the reaction amplitudes for those three sectors, 
1) nuclear elastic scattering, 2) an electromagnetic probe, and 
3) an electroweak probe for nuclear reactions, is systematically carried out
by introducing external vector (or minimally coupled photon) 
and axial-vector fields, which preserve
symmetry requirement for constructing nuclear reaction amplitudes.
In addition, a part of the reaction amplitudes, a dressed $^{16}$O propagator,
is shared by those three reactions, which can easily be identified by 
drawing Feynman diagrams. 
Once parameters for the dressed $^{16}$O propagator
are fitted to the elastic scattering data, it can be used for making nuclear
reaction amplitudes for the radiative $\alpha$ capture on $^{12}$C and the 
$\beta$ delayed $\alpha$ emission from $^{16}$N. 
Additional coupling constants appearing in the reaction amplitudes are fixed
by using experimental data for the corresponding reactions, and the $S_{E1}$
factor is extrapolated to stellar energies. 
Reviewing those calculations, as an example of application of EFTs to 
nuclear reactions, one may see the reliability of the present approach. 

This paper is organized as follows.
In Sec.~\ref{sec:2}, a short overview of EFTs for the present work is presented.
In Sec.~\ref{sec:3}, a construction of an EFT for the radiative $\alpha$ 
capture on $^{12}$C at stellar energy is discussed; 
counting rules of the reaction are mentioned,
and an effective Lagrangian is displayed.
In Sec.~\ref{sec:4}, an application of the EFT to the study of the elastic 
$\alpha$-$^{12}$C scattering at low energies is discussed, 
and in Sec.~\ref{sec:5}, the EFT is applied to the study of 
the $E1$ transition of radiative $\alpha$ capture on $^{12}$C. 
In Sec.~\ref{sec:6}, an application of the EFT to the $\beta$
delayed $\alpha$ emission from $^{16}$N is explored.
In Sec.~\ref{sec:7}, results and discussion of the present work are
summarized. In Appendix, the propagators and the vertex functions
for the elastic $\alpha$-$^{12}$C scattering are displayed.

\section{EFTs for nuclear reactions at low energies}
\label{sec:2}

EFTs are now a popular method for the studies of hadron and nuclear physics.
In this section, we review the progress of EFTs 
as regards to their application to nuclear reactions at low energies. 

The most representative example of an EFT 
is chiral perturbation theory ($\chi PT$),
a low energy effective field theory
of QCD~\cite{w-pa79,gl-ap84,gl-npb85,dgh-92}:
Hadrons are composite particles consisting of quarks and gluons,
which are described by $SU(3)$ color gauge theory.
The quarks and gluons, however, are confined
in the hadrons and never appear as free particles.
Meanwhile, the QCD Lagrangian approximately has
a global $SU(3)_R\times SU(3)_L$ flavor symmetry 
because of the light $uds$ quark masses.
The symmetry is spontaneously broken down to 
$SU(3)_V$ 
involving eight massless Goldstone bosons
in the ground state of QCD.
Because of the nature of the Goldstone bosons, its interactions vanish
in the zero momentum and zero light quark mass limits, and
one can expand a reaction amplitude perturbatively
in powers of the number of derivative and/or light meson mass factor
around the vanishing interaction limit.
One may notice that the picture of a hadron is completely altered
between QCD and $\chi$PT; a hadronic system, for example, a proton is
described by QCD as a many body system consisting of strongly interacting
quarks and gluons and by $\chi$PT 
as a heavy core surrounded by a cloud of the massless Goldstone bosons
in the chiral limit.

Weinberg suggested the first application of $\chi$PT to nuclear
physics~\cite{w-plb90,w-npb91}:
Because of an enhancement effect from two nucleon propagation,
which can alter the counting rules,
one may construct a nuclear potential according to the chiral order
counting rules 
from two-nucleon irreducible diagrams
in the time ordered perturbation theory.
To obtain a reaction amplitude,
one solves the Lippmann-Schwinger equation with the chiral potential. 
This approach matches to the traditional view of the nuclear
potential; the long range part of the nuclear potential consists of
the one-pion exchange, the intermediate range does of the two-pion exchange,
and the short range part is parameterized by models and has a hard repulsion
core. This picture can be reproduced order by order
in the Weinberg's counting scheme; at leading order (LO)
the nuclear potential consists of the two nucleon contact interactions
and the one-pion exchange contribution, at next-to leading order (NLO)
one includes the two pion exchange contributions and 
contact interactions with two derivatives or two meson mass factors, 
and so on.
However, because of a singular interaction due to the tensor force,
modification of the Weinberg's counting rules is necessary in order to 
make the scattering amplitudes cutoff-independent~\cite{ntvk-prc05}.

Instead of a perturbative expansion of the nuclear potential,
Kaplan, Savage and Wise(KSW) suggested to expand the two-nucleon
scattering amplitudes~\cite{ksw-plb98,ksw-npb98}.
An elusive point to construct a perturbation theory for the two-nucleon 
systems is how to treat unnaturally small scales appearing 
in the $s$-wave states,
the scattering length for $^1S_0$ channel, $a_{np}\simeq -23.7$~fm,
and the deuteron binding
energy for $^3S_1$ channel, $B_d\simeq 2.22$~MeV, 
compared to the typical scale of $\chi$PT, the pion mass,
$m_\pi\simeq140$~MeV.
To take account of those small scales in the $s$-wave two-nucleon 
scattering amplitudes, the LO
two-nucleon contact interactions are resummed up to infinite order while
the one-pion exchange is perturbatively included.~\footnote{
There are other approaches, e.g., which treat the one-pion-exchange piece 
non-perturbatively while the two-pion-exchange is treated perturbatively.
See, e.g., Ref.~\cite{ly-prc12}.
}
This scheme is known as the KSW counting scheme.
For studies of low energy reactions,
the theory is easily simplified as a pionless EFT
in which the pions are regarded as irrelevant degrees of freedom at high energy
and integrated out of the theory~\cite{crs-npa99}.

The pionless EFT is subsequently applied to the study of three nucleon systems.
Bedaque, Hammer and van Kolck studied the triton system 
by using the pionless theory~\cite{bhvk-npa00}~\footnote{
$^3$He system in the pionless EFT, 
including the Coulomb interaction, 
is studied 
in Ref.~\cite{ab-jpg10}.}.
They found a cyclic singularity, the so-called limit cycle, 
in the triton channel when scaling 
a momentum cutoff introduced in the coupled integral equations.
Along with the limit cycle, a universal feature 
emerges in the three-body system,
known as the Efimov effect~\cite{e-sjnp71,e-sjnp79}:
an infinite number of three-body bound states, whose binding energies appear
as a geometrical series, are accumulated 
at the threshold.
The Efimov sates in the triton system appear in a theoretical limit,
the so-called unitary limit, where the $s$-wave scattering lengths 
become infinite and the two-body binding energy vanish in the 
two-body propagators.
Thus, one can make a perturbative expansion around the unitary limit, 
which appears in the inverse of the two-body propagators.
It actually coincides 
with the well-known effective range expansion~\cite{b-pr49}.\footnote{
A dibaryon field, which has baryon number 2, is introduced
by Kaplan~\cite{k-npb97},
and it is useful to implement the effective range expansion
in a theory without pions~\cite{bs-npa01,ah-prc05}
and with perturbative pions~\cite{st-prc10,ah-prc12}.
}
\footnote{
Recently, a study for nuclear few-body systems around the unitary limit
was reported by K\"onig \textit{et al.}~\cite{ketal-prl17}.
}
It is conjectured that this vanishing point appears in QCD
at a slightly large unphysical pion mass, $m_\pi$ $\simeq$ 200~MeV, 
(see, e.g., Figs. 11 and 12 in Ref.~\cite{emg-npa03}).
Such an infrared point as a function of the pion mass 
for the triton system is  
studied by Braaten and Hammer~\cite{bh-prl03}.
To renormalize the cyclic singularity, one needs to promote the 
three-body contact interaction at LO. 
For the study of the triton system, 
one can fix the coupling constant of the three-body contact interaction
by using the triton binding energy.
In addition, one can  
choose a triton wavefunction as a LO constituent of an amplitude
and make a perturbative expansion around 
it~\cite{j-12,jp-fbs13,v-prc13,v-prc17}.

Along with various theoretical developments of EFTs
(a part of which we have briefly discussed above),
EFTs are also employed for phenomenological studies at low energies,
for which error estimates are important.
An error estimate for a reaction can be controlled
by using the perturbative expansion scheme of EFTs.
EFTs have been employed for the phenomenological studies of,
for example,
neutron $\beta$-decay~\cite{aetal-plb04,amt-plb09},
solar neutrino reactions on the deuteron~\cite{bck-prc01,aetal-plb03,ash-20},
radiative neutron capture on a proton
at BBN energies~\cite{r-npa00,achh-prc06},
proton-proton fusion~\cite{kr-npa99,bc-plb01,ashhk-plb08,cly-plb13},
$^3$He($\alpha$,$\gamma$)$^7$Be~\cite{hrv-epja18,znp-18},
and $^7$Be($p$,$\gamma$)$^7$B~\cite{znp-prc14,rfhp-epja14} in the Sun.
In the following sections, we discuss 
an application of an EFT to nuclear reactions for heavier nuclei
at low energies.

\section{EFT for radiative $\alpha$ capture on $^{12}$C at $T_G$}
\label{sec:3}

In this section, we discuss a construction of an EFT for the study of
radiative $\alpha$ capture on $^{12}$C at $T_G=0.3$~MeV. 
We first briefly review previous studies
and then construct an EFT for this reaction. 
In the following subsections, 
we analyze the typical and high energy-momentum scales for 
the radiative $\alpha$ capture reaction at $T_G$,
we also discuss which are 
the relevant and irrelevant physical degrees of freedom,
and from those we derive  
the power counting rules for our EFT description.
We also write down the effective Lagrangian for three reactions: 
elastic $\alpha$-$^{12}$C scattering, 
$E1$ transition of the radiative $\alpha$ capture on $^{12}$C, 
and $\beta$ delayed $\alpha$ emission from $^{16}$N.

\subsection{Introduction to the radiative $\alpha$ capture on $^{12}$C}

The radiative $\alpha$ capture on $^{12}$C,
${}^{12}$C($\alpha$, $\gamma$)${}^{16}$O,
is one of the fundamental reactions in
nuclear astrophysics, which determines
the ratio ${}^{12}$C/${}^{16}$O produced in helium burning~\cite{f-rmp84}.
The reaction rate, equivalently the astrophysical $S$-factor, of
the process at the Gamow peak energy, $T_G=0.3$~MeV~\footnote{
See the footnote~\ref{footnote;T}.
}, however, cannot
experimentally be determined due to the Coulomb barrier.
A theoretical model is necessary to be employed 
in order to extrapolate the cross section down to $T_G$
by fitting model parameters to available experimental data
measured at a few MeV or larger.
During the last half century, a lot of experimental and theoretical
studies for the process have been carried out.
For reviews, see, e.g., Refs.~\cite{bb-npa06,detal-17,chk-epja15}
and references therein.

In constructing a model for the radiative capture process, 
one needs to take into account 
the excited states of ${}^{16}$O~\cite{bb-npa06},
particularly, two excited bound states for
$l^\pi_{n\mbox{-}th}=1_1^-$ and $2_1^+$
just below the $\alpha$-${}^{12}$C breakup threshold at
$T=-0.045$ and $-0.24$~MeV~\footnote{
The kinetic energy $T$ denotes that of the $\alpha$-${}^{12}$C system
in the center of mass frame.
\label{footnote;T}
}, respectively,
as well as $1_2^-$ and $2_2^+$ resonant (second excited) states at
$T=2.42$ and $2.68$~MeV, respectively.
Thus the capture reaction to the ground state of ${}^{16}$O
at $T_G$ is expected to be $E1$ and $E2$ transitions dominant
due to the subthreshold $1_1^-$ and $2_1^+$ states.
While the resonant $1_2^-$ and $2_2^+$ states play
a dominant role in the available
experimental data at low energies, typically $1\le T \le 3$~MeV.
A minor contribution to the $S$-factor appears 
from so-called cascade transitions in which the initial 
$\alpha$ and $^{12}$C form an excited state of $^{16}$O 
emitting a photon, and 
it subsequently decays to the ground state of $^{16}$O.
Experimental data pertaining to processes for nuclear astrophysics
are compiled,
known as NACRE-II compilation~\cite{xetal-npa13},
in which the $S$-factor of
the ${}^{12}$C($\alpha$,$\gamma$)${}^{16}$O reaction
is estimated employing a potential model,
and reported uncertainty of the process is less than 20~\%.

Theoretical frameworks employed for the study
are categorized mainly into two~\cite{detal-17}:
the cluster models using generalized coordinate method~\cite{dbh-npa84}
or potential model~\cite{lk-npa85}
and the phenomenological models using the parameterization
of Breit-Wigner, $R$-matrix~\cite{lt-rmp58},
or $K$-matrix \cite{hdz-npa76}.
A recent trend of the study is to rely on intensive numerical analysis,
in which a larger amount of the experimental data relevant
to the study are accumulated, 
such as those from $^{12}$C($\alpha$,$\gamma$)$^{16}$O,
radiative proton capture on $^{15}$N to the ground state of $^{16}$O,
$^{15}$N($p$,$\gamma_0$)$^{16}$O,
$\beta$ delayed $\alpha$ emission from $^{16}$N, 
$^{16}$N($\beta\alpha$)$^{12}$C,
elastic $\alpha$-$^{12}$C scattering,
$^{12}$C($\alpha$,$\alpha$)$^{12}$C, 
and $\alpha$ transfer reactions, 
$^{12}$C($^6$Li,$d$)$^{16}$O,
$^{12}$C($^7$Li,$t$)$^{16}$O,
$^6$Li($^{12}$C,$d$)$^{16}$O, and 
$^7$Li($^{12}$C,$t$)$^{16}$O,
up to relatively high energies, $T \simeq 7$~MeV,
and a significant number of parameters of
the models are fitted to the data
by using computing power~\cite{detal-17,xetal-npa13,aetal-prc15}.
In the following, 
we discuss an alternative approach to estimate the $S$-factor at $T_G$; 
we discuss counting rules of the EFT at $T_G$ and display 
an effective Lagrangian for the reactions.

\subsection{
An EFT for the radiative $\alpha$ capture on $^{12}$C
}

In the study of the radiative $\alpha$ capture on $^{12}$C
at $T_G=0.3$~MeV
employing an EFT, at such a low energy,
one may regard the ground states of $\alpha$ and $^{12}$C
as point-like particles
whereas the first excited states of $\alpha$ and $^{12}$C
are chosen as irrelevant degrees of freedom, by which
a large momentum scale of the theory is determined.
An effective Lagrangian for the process is constructed
by using two spinless scalar fields for $\alpha$ and $^{12}$C,
and terms of the Lagrangian are expanded in terms of
the number of derivatives.
The expansion parameter of the theory
is $Q/\Lambda_H \sim 1/3$ where $Q$ denotes a typical
momentum scale $Q\sim k_G$: $k_G$ is the Gamow peak momentum,
$k_G = \sqrt{2\mu T_G}\simeq 41$~MeV, where
$\mu$ is the reduced mass of $\alpha$ and $^{12}$C.\footnote{
A typical length scale of the $\alpha$-$^{12}$C system is, thus,
$k_G^{-1}\simeq 4.8$~fm. This is comparable to the sum of the radii
of $\alpha$ and $^{12}$C obtained from a nuclear radius formula,
$r_A = 1.2 A^{1/3}$~fm; $r_\alpha + r_{^{12}C} = 1.2(4^{1/3}+12^{1/3})
= 4.65$~fm. 
Meanwhile, the emitted photon carries away almost all of the released energy,
$E_\gamma'\simeq 7.64$~MeV, and the length scale of the photon is larger
than the other length scales, $E_\gamma'^{-1}\simeq 26$~fm.
Thus, the photon may recognize the nuclear system as to be point-like.
}
$\Lambda_H$ denotes a large momentum scale
$\Lambda_H\simeq \sqrt{2\mu_4 T_{(4)}}$ or
$\sqrt{2\mu_{12} T_{(12)}}\sim 150$~MeV where
$\mu_4$ is the reduced mass of
one and three-nucleon system and $\mu_{12}$ is that of four and
eight-nucleon system. $T_{(4)}$ and $T_{(12)}$ are
the first excited energies
of $\alpha$ and $^{12}$C, respectively:
$T_{(4)}=20.21$~MeV for $0^+_2$ state of $^4$He
and $T_{(12)}=4.44$~MeV for $2^+_1$ state of $^{12}$C.
By including the terms up to next-to-next-to-leading order (NNLO),
for example,
one may obtain about 10\% theoretical uncertainty
for the process.

Coefficients appearing in an effective Lagrangian are fixed 
by using the experimental data which are measured at significantly 
higher energies than $T_G$. 
In the following sections, we employ the data from three processes,
elastic $\alpha$-$^{12}$C scattering measured at $T\simeq 2-5$~MeV,
$S_{E1}$ factor for $^{12}$C($\alpha$,$\gamma$)$^{16}$O at $T\simeq 1-3$~MeV,
and $\beta$ delayed $\alpha$ emission from $^{16}$N at $T\simeq 0.8-3.2$~MeV.
Thus, the perturbative scheme of the theory may not be reliable
at the energies where those experimental data are measured.
We first fix some parameters of the dressed $^{16}$O propagators,
parameterized in terms of effective range expansion,
by using the binding energies of the excited $^{16}$O states and 
fit the other parameters of the propagators to the phase shift data
of the elastic scattering measured at $T\simeq 2-5$~MeV. 
Because of non-perturbative nature of a
propagator we treat it as a non-perturbative quantity; the dressed 
$^{16}$O propagator commonly appears in the three processes, and
we keep it as a non-perturbative one and expand reaction amplitudes 
perturbatively around it. 
For the radiative capture process,
two additional parameters for the $E1$ transition 
amplitude of the radiative $\alpha$ capture process are fitted to the $S_{E1}$
factor data measured at $T\simeq 1-3$~MeV, 
and a value of the $S_{E1}$ factor is estimated at $T_G=0.3$~MeV.

For the $\beta$ delayed $\alpha$ emission from $^{16}$N,
parameters of the decay amplitudes are fitted 
to the experimental data measured at $T\simeq 0.8-3.2$~MeV. 
This study would explore a validity of the present approach.
One reason is that 
the $\beta$ delayed $\alpha$ emission data
are covered with the small energy region compared to that for the 
elastic scattering data. 
The other reason is, as mentioned before, to test 
a different parameterization for the dressed $^{16}$O propagator 
compared to that in the $R$-matrix or $K$-matrix analysis. 
In the conventional $R$-matrix analysis, 
the subthreshold $1_1^-$ state and the broad resonant $1_2^-$ state 
of $^{16}$O are represented by the Breit-Wigner formula 
and are linearly combined in the reaction matrix along
with a background contribution; a secondary maximum of the 
$\beta$ delayed $\alpha$ emission data
is known to be important to constrain an interference pattern among those 
levels~\cite{bd-npa88,jfhk-prc90,hfk-prc91}.
In the present approach, 
the interference between the $1^-$ subthreshold and resonant states doesn't 
exist because the subthreshold $1_1^-$ state and 
the broad resonant $1_2^-$ state of $^{16}$O is represented 
by a single dressed $^{16}$O propagator
(in terms of effective range expansion);
instead, we will see that 
the secondary peak can be reproduced by an interference between the amplitudes 
from a non-pole diagram and a pole diagram. 

An effective Lagrangian for the present study
is written as~\cite{a-epja16,a-18}
\bea
{\cal L} = {\cal L}_{ES} 
+ {\cal L}_{RC(+)}
+ {\cal L}_{BD(+)}
\,,
\eea
where ${\cal L}_{ES}$ is the Lagrangian for the elastic scattering 
process~\cite{a-epja16},
${\cal L}_{RC(+)}$ is for additional terms
for the radiative $\alpha$ capture process~\cite{a-18},
and
${\cal L}_{BD(+)}$ is for additional terms
for the $\beta$-delayed $\alpha$ emission from $^{16}$N. 

The Lagrangian ${\cal L}_{ES}$ may be written by using the composite $^{16}$O
fields consisting of $\alpha$ and $^{12}$C 
as~\cite{a-epja16,a-jkps18,bs-npa01,ah-prc05}
\bea
\lefteqn{{\cal L}_{ES} = \phi_\alpha^\dagger \left(
iD_0
+\frac{\vec{D}^2}{2m_\alpha}
+ \cdots
\right) \phi_\alpha
}
\nnb \\ &&
+ \phi_C^\dagger\left(
iD_0
+ \frac{\vec{D}^2}{2m_C}
+\cdots
\right)\phi_C
\nnb \\ && +
\sum_{l,n}C_{n}^{(l)}d_{(l)}^\dagger \left[
iD_0
+ \frac{\vec{D}^2}{2(m_\alpha+m_C)}
\right]^n d_{(l)}
\nnb \\ &&
- \sum_{l=0}^3y_{(l)}\left[
(\phi_\alpha O_l \phi_C)^\dagger d_{(l)}
+ d_{(l)}^\dagger(\phi_\alpha O_l \phi_C)
\right]
\nnb \\ &&
+ \cdots\,,
\label{eq;Lagrangian_SC}
\eea
where $\phi_\alpha$ ($m_\alpha$) and
$\phi_C$ ($m_C$) are fields (masses) of $\alpha$ and $^{12}$C,
respectively.
$D^\mu$ is a covariant derivative, 
$D^\mu = \partial^\mu + i{\cal Q}A^\mu$ where ${\cal Q}$ is a charge operator
and $A^\mu$ is the photon field.
The dots denote higher order terms.
$d_{(l)}$ represent $\alpha$ and $^{12}$C composite fields 
of angular momentum $l=0,1,2,3$:
$d_{(0)}$ for $l=0$, $d_{(1)i}$ for $l=1$,
$d_{(2)ij}$ for $l=2$, and $d_{(3)ijk}$ for $l=3$ 
where those spin states are represented by the subscripts $i$,
$ij$, and $ijk$ as Cartesian tensors~\cite{j-prb70}.
(We have suppressed those indices for the Cartesian tensors 
in the Lagrangian in Eq.~(\ref{eq;Lagrangian_SC}).)
Those composite fields are introduced to make an 
expansion around the unitary limit.
$C_{n}^{(l)}$ are coupling constants for
the propagation of the $\alpha$-$^{12}$C composite fields
for the $l$ channels,
and are related to effective range parameters
along with common multiplicative factors $1/y_{(l)}^2$.
As we will discuss later,
because of a modification of the counting rules 
for the elastic $\alpha$-$^{12}$C scattering
we include the terms up to $n=3$ for $l=0,1,2$
and those up to $n=4$ for $l=3$.
$O_l$ are projection operators
by which the $\alpha$-$^{12}$C system is projected to the $l$-th
partial wave states. Thus one has
\bea
&&
O_0 = 1\,,
\ \ \
O_{1,i} =
\frac{i\stackrel{\leftrightarrow}{D}_i}{M}
\equiv
i \left(
\frac{\stackrel{\rightarrow}{D}_C}{m_C} -
\frac{\stackrel{\leftarrow}{D}_\alpha}{m_\alpha}
\right)_i
\,,
\nnb \\ &&
O_{2,ij} =
- \frac{\stackrel{\leftrightarrow}{D}_i}{M}
\frac{\stackrel{\leftrightarrow}{D}_j}{M}
+ \frac13 \delta_{ij}
\frac{\stackrel{\leftrightarrow}{D}^2}{M^2}
\,,
\nnb \\ &&
O_{3,ijk} = 
-i
\frac{\stackrel{\leftrightarrow}{D}_i}{M}
\frac{\stackrel{\leftrightarrow}{D}_j}{M}
\frac{\stackrel{\leftrightarrow}{D}_k}{M}
\nnb \\ &&
\ \ \ 
+ i\frac15\left(
\delta_{ij}
\frac{\stackrel{\leftrightarrow}{D}_k}{M}
+\delta_{ik}
\frac{\stackrel{\leftrightarrow}{D}_j}{M}
+\delta_{jk}
\frac{\stackrel{\leftrightarrow}{D}_i}{M}
\right)
\frac{\stackrel{\leftrightarrow}{D}^2}{M^2}\,.
\eea

The Lagrangian ${\cal L}_{RC(+)}$ for additional terms to the study
for the $E1$ transition of the radiative $\alpha$ capture reaction
may be written down as~\cite{a-18}
\bea
{\cal L}_{RC(+)} &=&
- y_{}^{(0)}\left[
\phi_O^\dagger (\phi_\alpha
\phi_C)
+ (\phi_\alpha
\phi_C)^\dagger \phi_O
\right] 
\nnb \\ &&
-h^{(1)}\frac{y^{(0)}y_{(1)}}{\mu}\left[
(
{\cal O}_i^{(1)} \phi_O
)^\dagger
d_{(1)i} + \mbox{\rm H.c.}
\right]
+ \cdots
\label{eq;Lagrangian_RC}
\eea
with
\bea
{\cal O}_i^{(1)} = \frac{iD_i}{m_O}\,,
\eea
where $m_O$ is the mass of $^{16}$O in the ground state.
We note that 
because the $^{16}$O ground state appears only in the final state,
$\phi_O$ ($\phi_O^\dagger$) is introduced as a source field
for the $^{16}$O ground state in the final (initial) state.
In the first term in Eq.~(\ref{eq;Lagrangian_RC}),
for example, $\phi_O$ ($\phi_O^\dagger$)
destroys (creates) the $^{16}$O ground state, and
$(\phi_\alpha\phi_C)^\dagger$ [($\phi_\alpha \phi_C$)] fields
create (destroy) a $s$-wave $\alpha$-$^{12}$C state.
The transition rate between the $^{16}$O ground state and the
$s$-wave $\alpha$-$^{12}$C state is parameterized
by the coupling constant $y^{(0)}$, which is fixed
by using experimental data.
A contact interaction, the $h^{(1)}$ term,
is introduced to renormalize divergence
from loop diagrams for the radiative $\alpha$ capture reaction.

A weak decay amplitude is described in terms of the $V-A$ type 
current-current
interaction for semi-leptonic decay process, and its interaction
Hamiltonian density is given as
\bea
{\cal H}(x) = \frac{G_F}{\sqrt2} J_\mu^{(n)}(x)\cdot J^{(l)\mu}(x)\,,
\eea
where $G_F$ is the Fermi constant, $J^{(n)}_\mu(x)$ is a nuclear current
to be constructed by using the external vector and axial-vector fields
in the effective Lagrangian
while $J^{(l)\mu}(x)$ is a lepton current. In the energy-momentum
space it is given as
\bea
J^{(l)\mu}(q) =  \bar{u}_e(p')\gamma^\mu(1-\gamma_5)v_\nu(p)\,,
\eea
where $q$ is the momentum transfer, $q^\mu=p'^\mu-p^\mu$.

The Lagrangian ${\cal L}_{BD(+)}$ is for additional terms for the study
of the $\beta$ delayed $\alpha$ emission from $^{16}$N.
We construct ${\cal L}_{BD(+)}$ for interactions where
the initial $2_{1}^-$ ground state of $^{16}$N decaying
to the $1_1^-$ and $3_1^-$ states of $^{16}$O and
the $\alpha$-$^{12}$C breakup state for $l=1$ and 3 channels
through the Gamow-Teller transition, which
may be written as 
\bea
{\cal L}_{BD(+)} &=& 
C^{(l=1)}_a y_{(1)}
\left[ a_i \left(\phi_\alpha O_{1,j}\phi_C\right)
\right]^\dagger
\phi_{N,ij}
\nnb \\ &&
+C^{(l=1)}_b
\left(a_i d_{(1)j}\right)^\dagger
\phi_{N,ij}
\nnb \\ &&
+ C^{(l=3)}_a y_{(3)}
\left[\left(\phi_\alpha O_{3,ijk}\phi_C\right)a_k\right]^\dagger
\phi_{N,ij}
\nnb \\ &&
+ C^{(l=3)}_b
\left(d_{(3)ijk}a_k\right)^\dagger
\phi_{N,ij}
\nnb \\ &&
+ D^{(l=1)}_a y_{(1)}
\left[ a_i \left(\phi_\alpha O_1^2 O_{1,j}\phi_C\right)
\right]^\dagger
\phi_{N,ij}
\nnb \\ &&
+ D^{(l=1)}_b
\left[a_i ({\cal O}^{(1)2}d_{(1)j})\right]^\dagger
\phi_{N,ij}
\nnb \\ && + \cdots
\,,
\eea
where $a_i$ is the external axial-vector field, 
which generates an axial nuclear current for the Gamow-Teller transition,
and $\phi_{N,ij}$ is the source field for the initial $2_1^-$ ground state
of $^{16}$N.
$C_a^{(l=1)}$, $C_b^{(l=1)}$, $C_a^{(l=3)}$, and $C_b^{(l=3)}$ are
coefficients for $l=1$ and $l=3$ at LO, and 
$D_a^{(l=1)}$ and $D_b^{(l=1)}$ are those for $l=1$ at NNLO;
the indices of the squared operators, $O_1^2$ and 
${\cal O}^{(1)2}$, for the terms at NNLO 
are suppressed in the above equation.

\section{Elastic $\alpha$-$^{12}$C scattering in the cluster EFT}
\label{sec:4}

In this section, we construct dressed composite $^{16}$O propagators.
We first review the formalism of the effective range expansion for  
elastic $\alpha$-$^{12}$C scattering.
We then discuss a modification of the counting rules for a low
momentum expansion around the unitary limit, based on the observation 
in a comparison between experimental data and terms generated from 
the Coulomb self-energy. 
After including broad resonant states of $^{16}$O,
we fit the effective range parameters to phase shift data.
We then calculate asymptotic normalization coefficients (ANCs)
for the $1_1^-$ and $3_1^-$ states of $^{16}$O and compare them
to the previous results.  

\subsection{
Differential cross section of the elastic scattering
}

The differential cross section of the elastic $\alpha$-$^{12}$C scattering
(for two spin-0 charged particles)
in terms of the phase shifts is given by
(see, e.g., Ref.~\cite{lt-rmp58})
\bea
\sigma(\theta) &=&
\frac{d\sigma}{d\Omega} = |f(\theta)|^2
\nnb \\ &=&
\frac{1}{k^2} \left|
-\frac{\eta}{2\sin^2\frac12\theta}
\exp\left(-2i\eta \ln\sin\frac12\theta\right)
\right.  \nnb \\ && \left.
+ \frac12 i \sum_{l=0}^\infty (2l+1) \left[
\exp\left(2i\omega_l\right)
-U_l\right] P_l(\cos\theta)
\right|^2\,,
\label{eq;sigma-theta}
\eea
where $f(\theta)$ is the scattering amplitude
including both pure Coulomb part
and Coulomb modified strong interaction part,
$\theta$ is the scattering angle,
$k$ is the absolute relative momentum,
and
$\eta = \kappa/k$.
$\kappa$ is the inverse of the Bohr radius,
$\kappa = Z_\alpha Z_C\mu \alpha_E$ where $Z_\alpha$ and $Z_C$ are
the number of protons in $\alpha$ and $^{12}$C, respectively,
and $\alpha_E$ is the fine structure constant.
In addition,
$\omega_l$ is
the Coulomb scattering phase,
$\omega_l (= \sigma_l-\sigma_0) 
= \sum \arctan(\eta/s)$ for $s=1$ to $l$;
$\sigma_l$ are the Coulomb phase shifts,
$\sigma_l=\arg\Gamma(1+l+i\eta)$ with $l=0,1,2,\cdots$,
and
\bea
U_l = \exp\left[
2i(\delta_l + \omega_l)\right]\,.
\eea
$\delta_l$ are real scattering phase shifts.
$P_l(x)$ are the Legendre polynomials.
The scattering amplitudes are represented in terms of
$\delta_l$ as~\cite{lt-rmp58}~\footnote{
One can see the relation between $U_l$ and $A_l$ through 
the relation
\bea
\frac12 i\left[
\exp\left(2i\omega_l
\right) - U_l
\right]= 
\exp\left(2i\omega_l\right)\frac{1}{\cot\delta_l-i}\,.
\nnb
\eea
}
\bea
A_l &=& \frac{2\pi}{\mu}\frac{(2l+1)P_l(\cos\theta)e^{2i\sigma_l}}{
k\cot \delta_l-ik}\,.
\label{eq;Al-delta}
\eea

\begin{figure}
\begin{center}
\resizebox{0.5\textwidth}{!}{
  \includegraphics{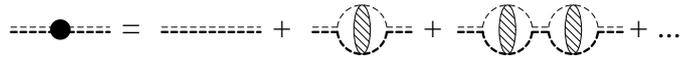}
}
\caption{
Diagrams for dressed $^{16}$O propagator.
A thick (thin) dashed line represents a propagator of $^{12}$C ($\alpha$),
and a thick and thin double dashed line with and without a filled circle
represent a dressed and bare $^{16}$O propagator, respectively.
A shaded blob represents
a set of diagrams consisting of all possible one-potential-photon-exchange
diagrams up to infinite order and no potential-photon-exchange one.
}
\label{fig;propagator}       
\end{center}
\end{figure}
\begin{figure}
\begin{center}
\resizebox{0.15\textwidth}{!}{%
  \includegraphics{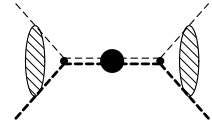}
}
\caption{
Diagram of the scattering amplitude.
See the caption of Fig.~\ref{fig;propagator} as well.
}
\label{fig;scattering_amplitude}      
\end{center}
\end{figure}
In the EFT,
the amplitudes of the elastic scattering
are calculated from diagrams depicted in
Figs.~\ref{fig;propagator} and \ref{fig;scattering_amplitude}.
We obtain the scattering amplitudes for $l$-th partial wave states
as~\cite{a-epja16,aetal-prc07,a-epja07}
\bea
A_l &=& \frac{2\pi}{\mu}\frac{(2l+1)P_l(\cos\theta) e^{2i\sigma_l}
W_l(\eta)
C_\eta^2}{
K_l(k) - 2\kappa H_l(k)
}\,,
\label{eq;Al}
\eea
with
\bea
&&
C_\eta^2 = \frac{2\pi\eta}{e^{2\pi\eta}-1}\,,
\ \ \
W_l(\eta) = \frac{\kappa^{2l}}{(l!)^2}
\prod^{l}_{n=0}\left(
1+\frac{n^2}{\eta^2}
\right)\,,
\nnb \\ &&
H_l(k) = W_l(\eta) H(\eta)\,,
\label{eq;Hl}
\eea
and
\bea
H(\eta) = \psi(i\eta) + \frac{1}{2i\eta} -\ln(i\eta)\,,
\label{eq;H}
\eea
where $\psi(z)$ is the digamma function.
Note that the function, $-2\kappa H_l(k)$, 
the Coulomb self-energy term, in the denominator
of the amplitude is obtained from the Coulomb bubble diagram
for the dressed propagator of $^{16}$O
in Fig.~\ref{fig;propagator}, and the factor,
$e^{2i\sigma_l}W_l(\eta)C_\eta^2$, in the numerator
is from the initial and final state Coulomb interactions
between $\alpha$ and $^{12}$C
in Fig.~\ref{fig;scattering_amplitude}.
In Appendix A, we display the renormalized dressed three-point
vertices for the initial and final Coulomb interactions
and the renormalized dressed composite $^{16}$O propagators 
for $l=0,1,2,3$.

The function $K_l(k)$ represents the interaction
due to the short range nuclear force
(compared with the long range Coulomb force),
which is obtained
in terms of the effective range parameters as
\bea
K_l(k) = -\frac{1}{a_l} + \frac12 r_l k^2 - \frac14 P_l k^4
+ Q_l k^6 - R_l k^8
+ \cdots\,.
\label{eq;Kl}
\eea
One can find that the expression obtained in Eq.~(\ref{eq;Al})
reproduces well the previous results reported
in Refs.~\cite{klh-jpg13,hot-npb73,h-jmp77,scb-prc10}.

By comparing the two expressions of the amplitudes $A_l$
in Eqs.~(\ref{eq;Al-delta}) and (\ref{eq;Al}),
one has a relation between the phase shift
and the effective range parameters in the denominator
of the scattering amplitudes, $D_l(k)$, as 
\bea
W_l(\eta)C_\eta^2 k\cot\delta_l = Re D_l(k)\,,
\label{eq;dell}
\eea
where 
\bea
D_l(k) = K_l(k) - 2\kappa H_l(k)\,.
\eea 

To estimate the ANC, $|C_b|$,
for the $0_2^+$, $1_1^-$, $2_1^+$, $3_1^-$ states of $^{16}$O,
we employ the definition of $|C_b|$~\cite{ir-prc84},
\bea
|C_b| &=&
\gamma_l^l\frac{\Gamma(l+1+|\eta_b|)} {l!}
\left(\left|-\frac{dD_l(k)}{dk^2}\right|_{k^2=-\gamma_l^2}\right)^{-\frac12}
\,,
\label{eq;Cb}
\eea
where $\eta_b=\kappa/k_b$;
$k_b = i\gamma_l$,
and $\gamma_l$ are binding momenta which will be given in the next subsection.

\subsection{
Effective range expansion and modification of the counting rules
}

We now discuss a modification of the counting rules~\cite{a-prc18},
based on an observation in Eq.~(\ref{eq;dell}).
The term on the left-hand-side of Eq.~(\ref{eq;dell}) is significantly
suppressed by the factor $C_\eta^2$ at low energies. Meanwhile, the Coulomb 
self-energy term, $-2\kappa H_l(k)$, on the right-hand-side of 
Eq.~(\ref{eq;dell}) turns out to be two orders of magnitude larger than the
term on the left-hand-side of the equation. 
Thus, we introduce terms in $K_l(k)$ as counter terms at LO,
by which those unnaturally large terms from the Coulomb 
self-energy term are subtracted.
We will discuss it in detail below. 

The effective range parameters in $K_l(k)$
are expanded in powers of $k^2$ whereas
the real part of the function $H_l(k)$
can be expanded in powers of $k^2$ as well.
For the function $H(\eta)$ in $H_l(k)$, one has
\bea
Re H(\eta) &=&
\frac{1}{12\kappa^2}k^2
+ \frac{1}{120\kappa^4}k^4
+ \frac{1}{252\kappa^6}k^6
+ \frac{1}{240\kappa^8}k^8
\nnb \\ &&
+ \cdots\,,
\eea
where $\kappa$ is the inverse of the Bohr radius,
$\kappa\simeq 245$~MeV, and regarded as another large scale of the theory.
Thus, the right-hand-side of equation, $Re D_l(k)$, in Eq.~(\ref{eq;dell})
can be expanded as a power series of $k^2$ for
both $K_l(k)$ and $2\kappa Re H_l(k)$.
Meanwhile, the left-hand-side
of Eq.~(\ref{eq;dell}) is suppressed by the factor $C_\eta^2$,
due to the Gamow factor $P=\exp(-2\pi\eta)$.

In the case of the $s$-wave, for example, the reported phase shift
at the smallest energy, $T_\alpha = 2.6$~MeV,
where $T_\alpha$ is the kinetic energy of $\alpha$ 
in the laboratory frame~\footnote{
The kinetic energies, $T$ in the center of mass frame 
and $T_\alpha$ in the laboratory frame,
are related by a relation $T=\frac34T_\alpha$. 
},
is $\delta_0=-1.893^\circ$~\cite{tetal-prc09}.
The factor $C_\eta^2$ becomes
$C_\eta^2\simeq 6 \times 10^{-6}$ at $k=104$~MeV
which corresponds to $T_\alpha=2.6$~MeV, and
the left-hand-side of Eq.~(\ref{eq;dell}) numerically becomes
$C_\eta^2 k\cot\delta_0 = -0.019$~MeV.
The function $2\kappa Re H_0(k)$ is expanded as
\bea
\lefteqn{2\kappa Re H_0(k) 
}
\nnb \\
&=&
\frac{1}{6\kappa}k^2
+ \frac{1}{60\kappa^3}k^4
+ \frac{1}{126\kappa^5}k^6
+ \frac{1}{120\kappa^7}k^8 + \cdots
\nnb \\ &=& 7.441 + 0.136 + 0.012 + 0.002 + \cdots \ \
\mbox{\rm (MeV)}\,,
\label{eq;S-wave}
\eea
at $k=104$~MeV. 
The numerical values in the third line
of Eq.~(\ref{eq;S-wave}) correspond to the terms appearing
in the second line of the equation in order.
One can see that the power series converges well, but
the first and second terms are two and one order of magnitude
larger than a value estimated by using the experimental data
in the left-hand-side of Eq.~(\ref{eq;dell}),
$-$0.019~MeV.
Those terms are unnaturally large, and thus it is
necessary to introduce a new renormalization method,
in which the counter terms remove the unnaturally large terms
and make the terms in a natural size.
In other words,
we assume that fitting polynomial functions
are represented as a natural power series at the low energy region,
and to maintain such polynomial functions,
large cancellations for the first and second terms
with the $r_l$ and $P_l$ effective terms, respectively, are expected.
So we include the three effective range parameters,
$r_l$, $P_l$, and $Q_l$, for the $l=0$ channel
as the counter terms.
The same tendency can be seen in the $l=1,2$ channels whereas
one needs to include four effective range parameters for the $l=3$ channel.
Thus, we employ the three effective range parameters,
$r_l$, $P_l$, $Q_l$ for the $l=0,1,2$ channels
and the four effective range parameters,
$r_l$, $P_l$, $Q_l$, $R_l$  for the $l=3$ channel
when fitting the parameters to the phase shift data.

At the binding energies of the excited 
$0_2^+$, $1_1^-$, $2_1^+$, $3_1^-$ states of $^{16}$O,
the amplitudes should have a pole
at $k_b = i\gamma_l$ where $\gamma_l$ are the binding momenta,
$\gamma_l = \sqrt{2\mu B_l}$;
$B_l$ denote the binding energies of $l_{i\textrm{-}th}^\pi$ excited states of $^{16}$O.
Thus the denominator of the scattering amplitude, $D_l(k)$,
should vanish at $k_b$.
Using this condition, a first effective range parameter, $a_l$,
is related to other effective range parameters as
\bea
-\frac{1}{a_l} &=&
\frac12 r_l\gamma_l^2
+ \frac14 P_l\gamma_l^4
+ Q_l \gamma_l^6
+ R_l \gamma_l^8
+ \cdots
\nnb \\ &&
+2\kappa H_l(k_b)\,,
\label{eq;moneoveral}
\eea
and the remaining effective range parameters are fixed by using the phase
shift data of the elastic scattering.\footnote{
In Ref.~\cite{a-jkps18},
we have also studied an inclusion of the ground $0_1^+$ state
of $^{16}$O in the parameter fitting 
for the elastic $\alpha$-$^{12}$C scattering at low energies. 
}

\subsection{
Numerical results for the elastic scattering
}

The effective range parameters, $r_l$, $P_l$, $Q_l$ for $l=0,1,2$ and 
$r_l$, $P_l$, $Q_l$, $R_l$ for $l=3$ are fitted, 
employing the standard $\chi^2$ fit~\footnote{
We employ a python package, {\tt emcee}\cite{fmetal-12},
for the fitting.
\label{footnote;emcee}
},
to the experimental 
phase shift data of the elastic $\alpha$-$^{12}$C scattering reported
by Tischhauser \textit{et al.}~\cite{tetal-prc09,epaps}.~\footnote{
Input data for the parameter fitting are the phase shifts of the
elastic $\alpha$-$^{12}$C scattering
for $l=0,1,2,3$~\cite{tetal-prc09,epaps},
which have been generated from the $R$-matrix analysis of the
elastic scattering data~\cite{tetal-prl02}.
In the input data files~\cite{epaps}, there are four column data:
the first column is the alpha energy, the second one the phase shift as
derived from the globalized Monte Carlo simulations, the third one is
the same phase shift randomized by the error from the Monte Carlo simulations,
and the fourth one is the error of the phase shifts from the Monte Carlo
simulations.
We have used the second column of the phase shift data in our previous
works~\cite{a-epja16,a-prc18} and the third column of the phase shift data
in our other works~\cite{a-jkps18,ya-19,a-18} for fitting.
}
The reported energy range of the data
is $2.6 \le T_\alpha \le 6.62$~MeV 
($2 \le T \le 5$~MeV.)~\cite{tetal-prc09}. 
Though we have discussed that a large energy scale of the theory 
is that for the first excited $2_1^+$ state of $^{12}$C, 
$T_{(12)}=4.44$~MeV ($T_{\alpha(12)} = 5.92$~MeV) in the previous section,
other large scales can emerge from resonate energies of $^{16}$O. 
As suggested by Teichmann~\cite{t-pr51},
below the resonance energies, the Breit-Wigner-type parameterization
for the resonances can be expanded in powers of the energy, and
one can obtain an expression of the amplitude in terms of
the effective range expansion.
Therefore, in the previous works~\cite{a-epja16,a-prc18},
we choose the energies of the resonant states,
$T_\alpha(0_3^+)=6.52$~MeV,
$T_\alpha(1_2^-)=3.23$~MeV,
$T_\alpha(2_2^+)=3.58$~MeV,
$T_\alpha(3_2^-)=5.92$~MeV,
as a large scale of the theory for each partial wave state.
After fitting the parameters to the data below the resonant energies,
we confirm large cancellations between the fitted parameters and 
those generated from the Coulomb self-energy terms,
e.g., in Eq.~(\ref{eq;S-wave}) while the perturbative series
of the effective range expansions at $T_G$ converges well
as expected by the theory. See Table V in Ref.~\cite{a-prc18}
and Table 2 in Ref.~\cite{a-jkps18}. 
The perturbative series, however, 
does not converge at the large, experimental energies. 
We will come back to the issue in the next section.

In our recent works~\cite{ya-19,a-18}, 
we include broad resonant states in the fitting 
because one may expect that because a relatively deep pole 
for a broad resonance state in the complex energy plane
is located at a distance from the real axis, it can possibly be represented 
by a polynomial function in terms of the effective range expansion.
We find that the broad resonant $1_2^-$ and $3_2^-$
states with the decay widths
$\Gamma(1_2^-) = 420\pm 20$~keV and
$\Gamma(3_2^-) = 800\pm 100$~keV
can be well incorporated into the fitting of the effective range parameters
whereas the narrow resonant $0_3^+$ and $2_2^+$ states with
the decay widths $\Gamma(0_3^+) = 1.5\pm 0.5$~keV and
$\Gamma(2_2^+) = 0.625\pm0.100$~keV cannot.
In addition, a tail-like structure from a higher resonant $1_3^-$ state at
$T_\alpha(1_3^-)=7.96$~MeV with the width $\Gamma(1_3^-) = 110\pm 30$~keV 
cannot be incorporated into the fitting for the high energy region of the data
at $T_\alpha\ge 6$~MeV for $l=1$ either.
Thus we use the maximum energies, $T_{\alpha,max}=5.5$,
6.0, 3.2, and 6.62~MeV for $l=0,1,2$, and 3, respectively,
of the data sets for the fitting where we assume that 
the resonant $2_1^+$ state of $^{12}$C, which could appear 
at $T_{\alpha(12)}=5.92$~MeV, would be negligible for $l=1$ and $l=3$.

\subsubsection{$l=0$ channel}

As discussed above, we use the phase shift data for $l=0$
up to the maximum energy $T_{\alpha,max} = 5.5$~MeV
due to the narrow resonant $0_3^+$ state
at $T_\alpha(0_3^+) = 6.52$~MeV with $\Gamma(0_3^+) = 1.5\pm 0.5$~keV.
To study the dependence on the choice of the data sets,
we employ four data sets up to different maximum energies,
$T_{\alpha, max} = 4.0$, 4.5, 5.0, and 5.5~MeV.

\begin{table*}
\begin{center}
\begin{tabular}{c|ccc|cc}\hline
$T_{\alpha,max}$ (MeV)  &
$r_0$ (fm) & $P_0$ (fm$^3$) & $Q_0$ (fm$^5$) &
        $a_0$ (fm) & $|C_b|$ (fm$^{-1/2}$)
\cr \hline
4.0 & $0.26851(5)$ & $-0.0354(33)$ & $0.0014(15)$ &
        $3.89\times 10^4$ & $4.5(39)\times 10^2$
\cr
4.5 & $0.26851(3)$ & $-0.0352(13)$ & $0.0015(5)$ &
        $4.84\times 10^4$ & $4.9(20)\times 10^2$
\cr
5.0 & $0.26849(3)$ & $-0.0358(8)$ & $0.0013(3)$ &
        $4.11\times 10^4$ & $4.1(8)\times 10^2$
\cr
5.5 & $0.26845(2)$ & $-0.0375(5)$ & $0.0006(2)$ &
        $3.10\times 10^4$ & $3.1(2)\times 10^2$
\cr \hline
\end{tabular}
\caption{
Center values and errors of the effective range parameters,
$r_0$, $P_0$, and $Q_0$, fitted to the data of the
data sets with the energy ranges,
$2.6~\mbox{\rm MeV}\le T_\alpha \le T_{\alpha,max}$
where $T_{\alpha,max}=4.0$, 4.5, 5.0, and 5.5~MeV.
Values of $a_0$ and those of $|C_b|$ for the $0_2^+$ state of $^{16}$O
are calculated by using the fitted $r_0$,  $P_0$, and $Q_0$ values.
For details, see the text.
\label{table:s-wave_parameters}
}
\end{center}
\end{table*}
In Table~\ref{table:s-wave_parameters},
fitted values and errors of the effective range parameters,
$r_0$, $P_0$, and $Q_0$, are displayed,
and calculated values of $a_0$ by using Eq.~(\ref{eq;moneoveral}) 
and those of the ANC, $|C_b|$, for the
$0_2^+$ state of $^{16}$O by using Eq.~(\ref{eq;Cb}) 
are also displayed.
$\chi^2/N$ and 
numbers of the data ($N$) 
for the fitting are 
$\chi^2/N(N)= 0.354(80)$, 0.393(149), 0.450(167), 0.552(230) for
$T_{\alpha,max}=4.0$, 4.5, 5.0, 5.5~MeV, respectively.
As seen in the table, the errors of those quantities become smaller as
the value of $T_{\alpha,max}$ increases (mainly because the number of
the data increases) while the center values of them are changing. 
The change of the center values of the coefficients may indicate
an effect of truncation error in the fitting 
and/or that of the narrow resonance.
If we include the data at high energies larger than $T_{\alpha,max}=5.5$~MeV
into the fitting, we have large $\chi^2/N$ values:
$\chi^2/N=1.36$ with those up to $T_{\alpha,max} = 6.0$~MeV,
and $\chi^2/N=11.3$ with those up to $T_{\alpha,max} = 6.6$~MeV.~\footnote{
In Ref.~\cite{a-20}, we recently studied an inclusion of the sharp 
resonant $0_3^+$ state of $^{16}$O and the first excited $2_1^+$ state 
of $^{12}$C in the study of elastic $\alpha$-$^{12}$C scattering 
for $l=0$ up to $T_{\alpha,max} = 6.62$~MeV.
We found that the $2_1^+$ state of $^{12}$C is redundant 
for fitting the phase shift data. 
To investigate its precise role, the inelastic open channel,
$\alpha +^{12}$C$^*(2_1^+)$, would be necessary to be included in the 
study of the elastic scattering above the excited energy of $^{12}$C. 
\label{footnote;l=0}
}

\subsubsection{$l=2$ channel}

For $l=2$, as discussed above, we cannot incorporate the narrow
resonant $2_2^+$ state at $T_\alpha(2_2^+)=3.58$~MeV with
$\Gamma(2_2^+) = 0.625\pm 0.100$~keV in the fitting.
Thus we choose the maximum energy of
the data set for the fitting as
$T_{\alpha,max} = 3.2$~MeV. Using the data of the data set we
fit effective range parameters as
\bea
r_2 &=& 0.15\pm 0.14\, \mbox{\rm fm$^{-3}$}\,,
\ \ \
P_2 = -1.2\pm 2.2\, \mbox{\rm fm$^{-1}$}\,,
\nnb \\  
Q_2 &=& 0.10\pm 0.93\, \mbox{\rm fm}\,.
\eea
We calculate a value of $a_2$ and that of the ANC, $|C_b|$, for the
subthreshold $2_1^+$ state of $^{16}$O as
$a_2 = 4.6\times 10^3$~fm$^5$ and $|C_b|=(3.1\pm 24.5)\times 10^4$~fm$^{-1/2}$,
respectively, where the number of the data is
$N=23$ and $\chi^2/N=0.22$.
One can see above that because the value of $\chi^2/N$ is quite small while
the error bars of those fitted quantities are large, we cannot deduce
a meaningful result for $l=2$ from the fitting.

\subsubsection{$l=1$ and $l=3$ channels}

For $l=1$, as discussed above, we can incorporate the broad resonant $1_2^-$
state at $T_\alpha(1_2^-)= 3.23$~MeV
with $\Gamma(1_2^-) = 420\pm 20$~keV
in the fitting for the effective range parameters
but cannot do a tail from the next broad resonant $1_3^-$ state at
$T_\alpha(1_3^-)=7.96$~MeV with $\Gamma(1_3^-) = 110\pm 30$~keV.
Thus the highest energy of the data for the fitting is
$T_{\alpha,max}=6.0$~MeV for $l=1$.
To study the dependence on the choice of the data sets,
we employ four data sets up to the different maximum energies,
$T_{\alpha, max} = 3.0$, 4.0, 5.0, and 6.0~MeV.

\begin{table*}
\begin{center}
\begin{tabular}{c|ccc|cc}\hline
$T_{\alpha,max}$ (MeV)  &
        $r_1$ (fm$^{-1}$) & $P_1$ (fm) & $Q_1$ (fm$^3$) &
        $a_1$ (fm$^3$) & $|C_b|$ (fm$^{-1/2}$)
\cr \hline
3.0 & $0.4157(9)$ & $-0.568(11)$ & $0.022(4)$ &
        $-1.316\times 10^5$ & $1.6(3)\times 10^{14}$
\cr
4.0 & $0.415266(49)$ & $-0.57481(56)$ & $0.02015(19)$ &
        $-1.665\times 10^5$ & $1.835(25)\times 10^{14}$
\cr
5.0 & $0.415272(20)$ & $-0.57474(20)$ & $0.02018(22)$ &
        $-1.658\times 10^5$ & $1.832(10)\times 10^{14}$
\cr
6.0 & $0.415273(9)$ & $-0.57473(9)$ & $0.02018(3)$ &
        $-1.658\times 10^5$ & $1.832(5)\times 10^{14}$
\cr \hline
\end{tabular}
\caption{
Center values and errors of the effective range parameters,
$r_1$, $P_1$ and $Q_1$, fitted to the data of the
data sets with the energy ranges,
$2.6~\mbox{\rm MeV} \le T_\alpha \le T_{\alpha,max}$
where $T_{\alpha, max} = 3.0$, 4.0, 5.0, and 6.0~MeV.
Values of $a_1$ and those of $|C_b|$ for the subthreshold
$1_1^-$ state of $^{16}$O are calculated by using
the fitted $r_1$, $P_1$, and $Q_1$ values.
For details, see the text.
\label{table:p-wave_parameters}
}
\end{center}
\end{table*}
In Table~\ref{table:p-wave_parameters},
fitted values and errors of the effective range parameters,
$r_1$, $P_1$, and $Q_1$ are displayed,
and calculated values of $a_1$ 
and those of the ANC, $|C_b|$, for the
$1_1^-$ state of $^{16}$O are displayed as well.
$\chi^2/N$ and numbers of the data ($N$)
for the fitting are
$\chi^2/N(N) = 0.872(13)$, 0.450(80), 0.509(167), 0.738(273)
for $T_{\alpha,max}=3.0$, 4.0, 5.0, 6.0, respectively.
As seen in the table, the errors of those quantities decrease as
$T_{\alpha,max}$ increases. It is worth pointing out that
the center values of them are almost not altered
even though $T_{\alpha,max}$ is changed.
Because the $\chi^2/N$ value for the data up to $T_{\alpha, max}=6.0$~MeV
is still smaller than 1, an effect from the resonant $2_1^+$ state of 
$^{12}$C is not significant for this channel.~\footnote{
Though the maximum energy of the data for fit is larger than the 
energy of the first excited $2_1^+$ state of $^{12}$C, 
$T_{(12)}=4.44$~MeV, there is no indication of a need to include
the $2_1^+$ state of $^{12}$C.
See the footnote \ref{footnote;l=0} as well. 
\label{footnote;l=1}
}
If we include the data at high energies larger than $T_{\alpha,max}=6.0$~MeV
into the fitting, we have a large $\chi^2/N$ value, $\chi^2/N=16.9$
with the data up to $T_{\alpha,max} = 6.6$~MeV.


For $l=3$, as discussed above, we can incorporate
the broad resonant $3_2^-$ state at $T_\alpha(3_2^-) = 5.92$~MeV with
$\Gamma(3_2^-) = 800\pm 100$~keV in the fitting of
the effective range parameters.
Thus we take available all energy range of the experimental
data for the parameter fitting; the highest energy of the data
is $T_{\alpha,max}=6.62$~MeV for $l=3$.
To study the choice of the data sets, we employ four data sets
up to different maximum energies,
$T_{\alpha, max} = 4.6$, 5.0, 6.0, and 6.62~MeV.

\begin{table*}
\begin{center}
\begin{tabular}{c|cccc|cc}\hline
$T_{\alpha,max}$ (MeV)  & $r_3$ (fm$^{-5}$) & $P_3$ (fm$^{-3}$) &
$Q_3$ (fm$^{-1}$) & $R_3$ (fm) & $a_3$ (fm$^7$) & $|C_b|$ (fm$^{-1/2}$)
\cr \hline
4.6 & $0.032(1)$ & $-0.50(12)$ & $0.28(9)$ & $-0.17(9)$ &
$-2.8\times 10^3$ & $2.3(82)\times 10^{2}$
\cr
5.0 & $0.0321(4)$ & $-0.507(57)$ & $0.276(42)$ & $-0.175(35)$ &
$-3.9\times 10^3$ & $4.3(265)\times 10^{2}$
\cr
6.0 & $0.0320(3)$ & $-0.494(11)$ & $0.285(6)$ & $-0.167(4)$ &
$-2.8\times 10^3$ & $2.2(7)\times 10^{2}$
\cr
6.62 & $0.0320(2)$ & $-0.495(6)$ & $0.285(3)$ & $-0.168(2)$ &
$-2.8\times 10^5$ & $2.3(4)\times 10^{2}$
\cr \hline
\end{tabular}
\caption{
Center vales and errors of
effective range parameters, $r_3$, $P_3$, $Q_3$, and $R_3$, fitted to the
data of the data sets with the energy ranges,
$2.6~\mbox{\rm MeV} \le T_\alpha \le T_{\alpha,max}$
where $T_{\alpha, max} = 4.6$, 5.0, 6.0, and 6.62~MeV.
Values of $a_3$ and the ANC, $|C_b|$,
for the $3_1^+$ state of $^{16}$O
are calculated by using the fitted $r_3$, $P_3$, $Q_3$, and $R_3$ values.
For details, see the text.
\label{table:f-wave_parameters}
}
\end{center}
\end{table*}
In Table~\ref{table:f-wave_parameters},
fitted values and errors of the effective range parameters,
$r_3$, $P_3$, $Q_3$, and $R_3$ are displayed,
and calculated values of $a_3$ 
and those of the ANC, $|C_b|$, for the
$3_1^-$ state of $^{16}$O are also displayed.
$\chi^2/N$ and numbers of the data ($N$) 
for the fitting are
$\chi^2/N(N)= 0.49(154)$, 0.45(167), 0.49(273), 0.50(354)
for $T_{\alpha,max}=4.6$, 5.0, 6.0, 6.62~MeV, respectively.
As seen in the table, the errors of those quantities decrease as
$T_{\alpha,max}$ increases. It is interesting to point out that,
as the same as that we have seen for $l=1$,
the center values of them are almost not altered
even though $T_{\alpha,max}$ is changed
except for the deviations of $a_3$ and $|C_b|$ and a large error of $|C_b|$
for $T_{\alpha,max}=5.0$~MeV compared to the corresponding values
for the other $T_{\alpha,max}$.
That may stem from negative correlations between the parameters
for $T_{\alpha,max}=4.6$~MeV
and almost no correlations between those for $T_{\alpha,max}=5.0$~MeV.
(One can find the correlations between the parameters in 
Table VI in Ref.~\cite{ya-19}.)
The effect from the resonant $2_1^+$ state of $^{12}$C is not significant
either because of the small $\chi^2/N$ value for the data up to 
$T_{\alpha,max}=6.62$~MeV.~\footnote{
See the footnote \ref{footnote;l=1}.
}

\subsubsection{Comparison of the ANCs}

\begin{table}
\begin{center}
\begin{tabular}{c|cc}\hline
 & $1_1^-$ & $3_1^-$ \cr \hline
$|C_b|$ (fm$^{-1/2}$)
 & $1.832(5)\times 10^{14}$ & $2.4(4)\times 10^2$
\cr \hline
\end{tabular}
\caption{
Our result of the ANCs for the $1_1^-$ and $3_1^-$ states of
$^{16}$O.
\label{table:Cbs}
}
\end{center}
\end{table}
In Table~\ref{table:Cbs},
we summarize our result of the ANCs for the $1_1^-$ and $3_1^-$
states of $^{16}$O where we show the results obtained from the largest
data sets for the fitting.
Because of the change of the center values and the large uncertainty 
for the ANCs for the $0_2^+$ and $2_1^+$ states 
of $^{16}$O, respectively, 
we do not include those results.
We suppress the comparison about the ANCs for the $0_2^+$ and 
$2_1^+$ states below.

The ANCs for the $1_1^-$ and $2_1^+$ states of $^{16}$O have intensively
been studied because the major contribution of the $S$-factor of the
radiative $\alpha$ capture on $^{12}$C at $T_G$ 
come from the $E1$ and $E2$ transitions
due to the subthreshold $1_1^-$ and $2_1^+$ states of $^{16}$O.
In our previous work~\cite{a-prc18},
we obtained $(1.6-1.9)\times 10^{14}$~fm$^{-1/2}$ for the $1_1^-$ state,
which agrees well with the result presented above.
Our result underestimates, by about 10\%, compared
to the other theoretical estimates:
$(2.22-2.24)\times 10^{14}$~fm$^{-1/2}$ obtained 
from a potential model calculation
by Katsuma~\cite{k-prc08},
and $2.14(6)\times 10^{14}$~fm$^{-1/2}$ and 
$2.073\times 10^{14}$~fm$^{-1/2}$ from
a new parameterization method
by Ramirez Suarez and Sparenberg~\cite{rss-prc17}
and by Orlov {\it et al.}~\cite{oetal-prc17}, respectively.
Our result, on the other hand,
also underestimates or agrees well with
the experimental results within the reported errors:
$(2.10\pm 0.14)\times 10^{14}$~fm$^{-1/2}$ obtained from
the $^6$Li($^{12}$C,$d$)$^{16}$O reaction
by Avila {\it et al.}~\cite{aetal-prl15},
$(2.00\pm 0.35)\times 10^{14}$~fm$^{-1/2}$ 
from the $^{12}$C($^7$Li,$t$)$^{16}$O
reaction by Oulebsir {\it et al.}~\cite{oetal-prc12}, and
$(2.08\pm 0.20)\times 10^{14}$~fm$^{-1/2}$ 
from the $^{12}$C($^6$Li,$d$)$^{16}$O and
$^{12}$C($^7$Li,$t$)$^{16}$O reactions
by Brune {\it et al.}~\cite{betal-prl99}.

Only some studies for the ANCs for the $0_2^+$ and $3_1^-$ states 
of $^{16}$O have been reported so far, though those ANCs are important
to fix the $S$-factor of the radiative $\alpha$ capture reaction at $T_G$
through the cascade transitions for the $R$-matrix analysis.  
Our result of the ANC of the $3_1^-$ state is large compared to that
in our previous result, $(1.2-1.5)\times 10^2$~fm$^{-1/2}$~\cite{a-prc18}.
This is because of an error of the numerical code in the previous study
and the use of the different data set of the phase shift for fitting.
Our result of the ANC for the $3_1^-$ state overestimates
an experimental result, $(1.39\pm 0.09)\times 10^2$~fm$^{-1/2}$ obtained by
Avila {\it et al.}~\cite{aetal-prl15}.

\section{
The $S_{E1}$-factor of radiative $\alpha$ capture on $^{12}$C 
}
\label{sec:5}

In this section, we discuss how to estimate the $S_{E1}$-factor of 
radiative $\alpha$ capture on $^{12}$C by using an EFT.
First, we explain the formalism to calculate the $S_{E1}$ and 
then we derive the $E1$ transition amplitudes up to NLO 
from the effective Lagrangian. 
We then discuss a suppression of the $E1$ transition amplitudes between
$I=0$ nuclear states, and, after fitting two parameters to the data,
we estimate the $S_{E1}$-factor at $T_G$.

\subsection{
$S_{E1}$ factor and radiative $\alpha$ capture amplitude
}

The $S_{E1}$-factor is represented by using
the total cross section $\sigma_{E1}$ from the $E1$ transition as
\bea
S_{E1}(T) &=& \sigma_{E1}(T)Te^{2\pi\eta}\,,
\eea
with
\bea
\sigma_{E1}(T) &=& \frac43\frac{\alpha_E \mu E_\gamma'}{p
(1+E_\gamma'/m_O)}
|X^{(l=1)}|^2\,,
\eea
where 
$T$ is the kinetic energy of the initial $\alpha$-$^{12}$C
state in the center of mass frame, 
$T=p^2/(2\mu)$; $p$ is the magnitude of relative 
momentum between $\alpha$ and $^{12}$C.
$E_\gamma'$ is the photon energy,
\bea
E_\gamma' \simeq B_0 + T - \frac{1}{2m_O}(B_0+T)^2\,,
\eea
where $B_0$ is the $\alpha$-$^{12}$C breakup energy from the ground
state of $^{16}$O; 
$B_0=m_O-m_\alpha-m_C=7.162$~MeV.
One may notice that $B_0$ is larger than the 
large energy scale of the theory, the first excited energy
of $^{12}$C, $T_{(12)}=4.44$~MeV. 
Because the released large energy is carried away by the outgoing photon, 
the final nuclear state remains in a state with a typical energy scale.
We will discuss how one can avoid invoking a resonant state originated
from the first excited $2_1^+$ state of $^{12}$C below.
$X^{(l=1)}$ is a transition amplitude which
will also be shown in the following. 

\begin{figure}
\begin{center}
\resizebox{0.5\textwidth}{!}{%
  \includegraphics{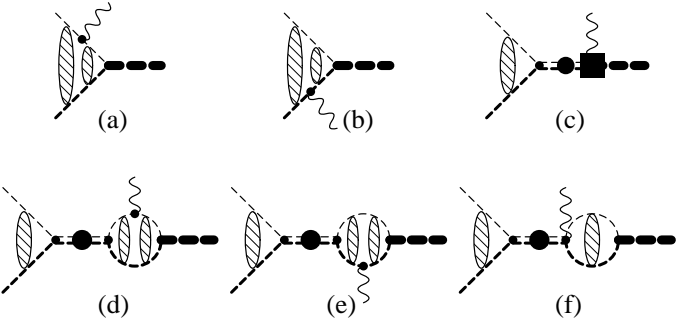}
}
\caption{
Diagrams for the radiative $\alpha$ capture process 
from the initial $p$-wave $\alpha$-$^{12}$C state.
A wavy line denotes the outgoing photon,
a thick and thin double dashed line with a filled circle in the
intermediate state,
whose diagrams are displayed in Fig.~\ref{fig;propagator},
the dressed composite $^{16}$O propagator for $l=1$,
and a thick dashed line in the final state
the ground ($0_1^+$) state of $^{16}$O.
See the caption of Fig.~\ref{fig;propagator} as well.
}
\label{fig;capture_amplitudes}       
\end{center}
\end{figure}

In Fig.~\ref{fig;capture_amplitudes},  diagrams of the radiative $\alpha$
capture process from the initial $\alpha$-$^{12}$C state for $l=1$ 
to the $^{16}$O ground ($0_1^+$)
state up to NLO are depicted,
in which the Coulomb interaction between
$\alpha$ and $^{12}$C is taken into account,
and 
in Fig.~\ref{fig;propagator},
those for dressed composite propagators of $^{16}$O consisting of $\alpha$
and $^{12}$C for $l=1$ are depicted. 
The propagator is obtained in the previous section.
(See Appendix as well.)

The radiative $\alpha$ capture amplitude for the initial $l=1$ state is
presented as
\bea
A^{(l=1)} &=& \vec{\epsilon}_{(\gamma)}^* \cdot \hat{p} X^{(l=1)}\,,
\eea
where $\vec{\epsilon}_{(\gamma)}^*$ is the polarization vector
of outgoing photon 
and $\hat{p}=\vec{p}/|\vec{p}|$; $\vec{p}$ is
the relative momentum of the initial $\alpha$ and $^{12}$C.
The amplitude $X^{(l=1)}$ is decomposed as
\bea
X^{(l=1)} &=&
X^{(l=1)}_{(a+b)}
+ X^{(l=1)}_{(c)}
+ X^{(l=1)}_{(d+e)}
+ X^{(l=1)}_{(f)}\,,
\eea
where those amplitudes correspond to the diagrams depicted in
Fig.~\ref{fig;capture_amplitudes}.

We follow the calculation method suggested by
Ryberg {\it et al.}~\cite{rfhp-prc14},
in which Coulomb Green's functions are represented in the coordinate space
satisfying appropriate boundary conditions.
Thus we obtain the expression of those amplitudes 
in the center of mass frame as
\bea
X_{(a+b)}^{(l=1)} &=&
2 y^{(0)}
e^{i\sigma_1} \Gamma(1+\kappa/\gamma_0)
\int_0^\infty drrW_{-\kappa/\gamma_0,\frac12}(2\gamma_0 r)
\nnb \\ && \times
\left[
\frac{Z_\alpha \mu}{m_\alpha}j_0\left(
\frac{\mu}{m_\alpha} k'r
\right)
-\frac{Z_C \mu}{m_C}j_0\left(
\frac{\mu}{m_C} k'r
\right)
\right]
\nnb \\ && \times
\left\{
\frac{\partial }{\partial r}\left[
\frac{F_1(\eta,pr)}{pr}
\right]
+2 \frac{F_1(\eta,pr)}{pr^2}
\right\}\,,
\label{eq;Xab}
\\
X_{(c)}^{(l=1)} &=& +
y^{(0)}
h^{(1)R}
\frac{6\pi Z_O}{\mu m_O}
\frac{e^{i\sigma_1}p\sqrt{1+\eta^2}C_\eta}{K_1(p) - 2\kappa H_1(p)}\,,
\label{eq;Xc}
\\
X_{(d+e)}^{(l=1)} &=&
+i
\frac{2}{3}y^{(0)}
\frac{
e^{i\sigma_1}p^2\sqrt{1+\eta^2}C_\eta
}{
K_1(p)-2\kappa H_1(p)
}
\Gamma(1+\kappa/\gamma_0)
\Gamma(2+i\eta)
\nnb \\ && \times
\int_{r_C}^\infty drrW_{-\kappa/\gamma_0,\frac12}(2\gamma_0 r)
\nnb \\ && \times
\left[
\frac{Z_\alpha \mu}{m_\alpha}j_0\left(
\frac{\mu}{m_\alpha} k'r
\right)
-\frac{Z_C \mu}{m_C}j_0\left(
\frac{\mu}{m_C} k'r
\right)
\right]
\nnb \\ && \times
\left\{
\frac{\partial }{\partial r}\left[
\frac{W_{-i\eta,\frac32}(-2ipr)}{r}
\right]
+2\frac{W_{-i\eta,\frac32}(-2ipr)}{r^2}
\right\}\,,
\nnb \\
\label{eq;Xde}
\\
X_{(f)}^{(l=1)} &=& -
3 y^{(0)}
\mu
\left[
-2\kappa H(\eta_{b0})
\right]
\left(
\frac{Z_\alpha}{m_\alpha} - \frac{Z_C}{m_C}
\right)
\nnb \\ && \times
\frac{e^{i\sigma_1}p\sqrt{1+\eta^2}C_\eta}{K_1(p) - 2\kappa H_1(p)}\,,
\label{eq;Xf}
\eea
where $k'$ is the magnitude of outgoing photon momentum, 
$Z_O$ is the number of protons in $^{16}$O,
and
$\gamma_0$ is the binding momentum of the ground state of $^{16}$O,
$\gamma_0=\sqrt{2\mu B_0}$.
$\Gamma(z)$ and $j_l(x)$ are gamma function and spherical Bessel function,
respectively,
while $F_l(\eta,\rho)$ and $W_{k,\mu}(z)$ are regular Coulomb function
and Whittaker function, respectively.
The functions, $H_1(p)$, $H(\eta)$ (where $\eta = \kappa/p$), and $K_1(p)$,
have been introduced in Eqs.~(\ref{eq;Hl}), (\ref{eq;H}),
and (\ref{eq;Kl}), respectively. 

Regarding the divergence from the loop integrals,
the loops of the diagrams (a) and (b)
in Fig.~\ref{fig;capture_amplitudes} are finite
while those of the diagrams (d) and (e) lead to a log divergence
in $X^{(l=1)}_{(d+e)}$ in the limit, $r\to 0$.
We introduce a short range cutoff $r_C$
in the $r$ integral in Eq.~(\ref{eq;Xde}),
and the divergence is renormalized by the counter term,
$h^{(1)}$.
The loop of the diagram (f) diverges and
is renormalized by the $h^{(1)}$ term as well.
Thus we have
\bea
h^{(1)R} = h^{(1)} -\mu
\frac{m_O}{Z_O}\left(
\frac{Z_\alpha}{m_\alpha}
-\frac{Z_C}{m_C}
\right)\left[
I_{(d+e)}^{div}
+ J_0^{div}
\right]\,,
\eea
where $I_{(d+e)}^{div}$ is the divergence term from the diagrams (d) and (e)
and $J_0^{div}$ is that from the diagram (f); we have
\bea
I_{(d+e)}^{div} &=& -\frac{\kappa\mu}{9\pi}\int_0^{r_C}\frac{dr}{r}\,,
\nnb \\ 
J_0^{div} &=& \frac{\kappa\mu}{2\pi}\left[
\frac{1}{\epsilon}
-3C_E
+ 2
+\ln\left(
\frac{\pi\mu_{DR}^2}{4\kappa^2}
\right)
\right]\,,
\eea
where, when we derive the expression of $J_0^{div}$,
the dimensional regularization in $4-2\epsilon$ space-time dimensions
is used; $C_E=0.577\cdots$ and $\mu_{DR}$ is a scale factor
from the dimensional regularization.
$h^{(1)R}$ is a renormalized coupling constant
which is fixed by experiment.

We have used the two regularization methods when calculating the
loop diagrams (d) and (e) and the loop diagram (f).
Some different regularization methods result in different expressions
for divergent terms and constant terms but the same expression for
functional terms (such as mass or momentum dependence terms)~\cite{r-npb96}.
Thus the different regularization methods may be adapted by 
adjusting a value of the coefficient $h^{(1)R}$ when we send a value of 
$r_C$ sufficiently small. 
Recently, Higa, Rupak, and Vaghani reported that the divergent terms
are exactly canceled with each other among those diagrams when they
calculate one-photon-exchange diagrams for 
the $\alpha_E$-order terms using 
the dimensional regularization~\cite{hrv-epja18}.

We now discuss how one can avoid invoking a resonant state 
due to the released energy from the radiative $\alpha$ capture reaction.
As mentioned above, the both initial and final nuclear states remain
in the states at the typical energies because the photon carries away almost 
all of the released energy. 
A large energy gap, then, appears in the intermediate state, i.e., in the loop
diagrams; the initial $p$-wave $\alpha$-$^{12}$C state is 
in a typical energy state while, after the photon is emitted, 
the energy gap occurs 
as a deep binging energy for the ground state of $^{16}$O
in the $s$-wave $\alpha$-$^{12}$C propagation. 
Because the binding energy for the ground state of $^{16}$O is far below 
the typical energies for the $\alpha$-$^{12}$C propagation, 
its physical effect will not be significant.
Moreover, the excited energy for 
the resonant $2_1^+$ state of $^{12}$C
is located at farther above than that for the 
$\alpha$-$^{12}$C breakup threshold, thus
an effect from the resonant $2_1^+$ state of $^{12}$C 
will hardly be seen. 
Meanwhile the loop integrals may pick up the deep momentum scale. 
From the loop diagram (f), for example, 
when the Coulomb interaction is ignored,
the large momentum scale $\gamma_0\simeq 200$~MeV is picked up
in the numerator of the amplitude.
It causes the emergence of a term which does not obey the counting
rules.~\footnote{
A method to renormalize a term which does not obey counting rules
in manifestly Lorentz invariant baryon chiral perturbation theory,
is known as the extended on mass shell (EOMS)
scheme~\cite{fgjs-prd03,af-prd07}. 
One can similarly renormalize the term proportional to $\gamma_0$ in the counter
term, $h^{(1)R}$, even when the Coulomb interaction does not exist.
}
In the present case,
the large momentum scale $\gamma_0$ from the ground state energy of $^{16}$O
appears as a ratio $\kappa/\gamma_0$,
due to the non-perturbative Coulomb interaction,
where $\kappa$ is another large momentum scale, $\kappa\simeq 245$~MeV.
The finite term $-2\kappa H(\eta_{b0})$
in $X_{(f)}^{(l=1)}$ from the loop of the diagram (f)
with $\eta_{b0}=\kappa/(i\gamma_0)$
is reduced to a typical momentum scale, $-2\kappa H(\eta_{b0}) = 25.8$~MeV.

\subsection{
Suppression of the $E1$ transition and mixture of isospin $I=1$ state
}

Before fitting the parameters to available experimental data,
we discuss three issues: non-perturbative treatment of the 
dressed $^{16}$O propagator for $l=1$, 
suppression of the $E1$ transition amplitude, 
and a mixture of isospin $I=1$ state.

For the radiative $\alpha$ capture amplitudes, 
whose expressions are displayed 
in Eqs~(\ref{eq;Xab}), (\ref{eq;Xc}), (\ref{eq;Xde}), (\ref{eq;Xf}),
we have two limits for perturbative expansion: The one appears
in the denominator of the transition amplitudes; the dressed $^{16}$O 
propagator is expanded around the unitary limit. 
The other appears in the numerator of the transition amplitudes as
loop and vertex corrections. 
As discussed above, the perturbative expansion in the 
denominator in terms of the effective range expansion is valid at $T_G$
and does not converge at the energies where the experimental data are 
available for fitting.
Meanwhile, because the phase shift data for $l=1$ are reproduced very well
(as we will see in Fig.~\ref{fig;delta1})
by means of the effective range expansion,
we treat the dressed $^{16}$O propagator as a non-perturbative quantity
and perturbatively expand the transition amplitudes around it.  

An order of an amplitude from each of the diagrams
is found by counting the number of momenta
of vertices and propagators in a Feynman diagram;
one has a LO amplitude from the diagram (c)
because the contact $\gamma$-$d_i$-$\phi_O$ vertex of the $h^{(1)R}$ term
does not have a momentum dependence, and
NLO amplitudes from the other diagrams in Fig.~\ref{fig;capture_amplitudes}.
One may notice a large suppression factor,
$Z_\alpha/m_\alpha - Z_C/m_C$, appearing in $X^{(l=1)}_{(f)}$;
$(m_O/Z_O)(Z_\alpha/m_\alpha - Z_C/m_C)\simeq -6.5\times 10^{-4}$.
Similar suppression effect can be found in $X^{(l=1)}_{(a+b)}$ and
$X^{(l=1)}_{(d+e)}$ as well; we denote those amplitudes as $X^-$,
and when changing the minus sign to the plus one
in the front of the spherical Bessel function $j_0(z)$
in Eqs.~(\ref{eq;Xab}) and (\ref{eq;Xde}), we do them as $X^+$.
We thus have
$|X^{(l=1)-}_{(a+b)}/X^{(l=1)+}_{(a+b)}| \simeq 8.7\times 10^{-4}$ and
$|X^{(l=1)-}_{(d+e)}/X^{(l=1)+}_{(d+e)}| \simeq 3.6\times 10^{-4}$ at the
energy range, $T=0.9-3$~MeV,
at which we fit the parameters to
the experimental $S_{E1}$ data in the next subsection.
The suppression effect is common among those amplitudes
from the diagrams (a), (b), (d), (e), (f) at NLO.
Thus, the radiative $\alpha$ capture rate will be well controlled 
by the coefficient $h^{(1)R}$ from the diagram (c) at LO.

The strong suppression effect mentioned above is well known;
the $E1$ transition is strongly suppressed between isospin-zero
($N=Z$) nuclei. This mechanism is recently reviewed and studied
for $\alpha(d,\gamma)^6$Li reaction by Baye and
Tursunov~\cite{bt-jpg18}.
In the standard microscopic calculations with the long-wavelength
approximation, the term proportional to $Z_1/m_1 - Z_2/m_2$ vanishes
because of the standard choice of mass of nuclei as $m_i = A_i m_N$
where $A_i$ is the mass number of $i$-th nucleus and $m_N$ is the
nucleon mass. We have strongly suppressed but non-zero contribution above
because of the use of the physical masses for $\alpha$ and $^{12}$C.
The small but non-vanishing $E1$ transition for the $N=Z$ cases
has intensively been studied in the microscopic calculations and
can be accounted by two effects:
The one is the second order term of the
$E1$ multipole operator in the long-wavelength approximation~\cite{db-plb83},
and the other is due to the mixture of the small $I=1$ configuration
in the actual nuclei~\cite{db-npa86}.
In the present approach, the first one may be difficult to incorporate
in the point-like particles while the second one could be introduced
from a contribution at high energy:
At $T\simeq 5$~MeV and 8.5~MeV above the $\alpha$-$^{12}$C breakup threshold,
$p$-$^{15}$N and $n$-$^{15}$O breakup channels, respectively, are open, and
$I=1$ resonant states of $^{16}$O start emerging
(along with the $I=1$ isobars, $^{16}$N, $^{16}$O, and $^{16}$F).
We might have introduced
the $p$-$^{15}$N and $n$-$^{15}$O fields
as relevant degrees of freedom in the theory.
The $p$-$^{15}$N and $n$-$^{15}$O fields, then, appear
in the intermediate states,
as $p$-$^{15}$N or $n$-$^{15}$O propagation,
in the loop diagrams (d), (e), (f)
in Fig.~\ref{fig;capture_amplitudes}
instead of the $\alpha$-$^{12}$C propagation.
One may introduce a mixture of the isospin $I=0$ and $I=1$ states
in the $p$-$^{15}$N or $n$-$^{15}$O propagation,
and the strong $E1$ suppression is circumvented in the loops.
(The contribution from the $p$-$^{15}$N and $n$-$^{15}$O channels
for the $^{12}$C($\alpha$,$\gamma$)$^{16}$O reaction
has already been studied in the microscopic approach~\cite{db-prc87}.)
In our work, however,
the $p$-$^{14}$N and $n$-$^{15}$O fields are regarded
as irrelevant degrees of freedom at high energy
and integrated out of the effective Lagrangian.
Its effect, thus, is embedded in the coefficient of the contact
interaction, the $h^{(1)R}$ term, in the diagram (c)
while the $h^{(1)R}$ term is fitted to the experimental $S_{E1}$ data
in the next subsection.

\subsection{
Numerical results for the radiative $\alpha$ capture reaction
}

We have five parameters 
in the radiative $\alpha$ capture amplitudes
to fit to the data;
three parameters, $r_1$, $P_1$, $Q_1$, are fitted to
the phase shift data of the elastic scattering, and the other two parameters,
$h^{(1)R}$ and $y^{(0)}$,
are to the experimental $S_{E1}$ data.
The standard $\chi^2$-fit is performed by employing
a Markov chain Monte Carlo method
for the parameter fitting.~\footnote{
See the footnote~\ref{footnote;emcee}.
}
The phase shift data for $l=1$ are taken from
Tischhauser {\it et al.}'s paper~\cite{tetal-prc09},
and the experimental $S_{E1}$ data are
from the literature
summarized in Tables V and VII in Ref.~\cite{detal-17}:
Dyer and Barnes (1974)~\cite{ex1},
Redder {\it et al.} (1987)~\cite{ex2},
Ouellet {\it et al.} (1996)~\cite{ex3},
Roters {\it et al.} (1999)~\cite{ex4},
Gialanella {\it et al.} (2001)~\cite{ex5},
Kunz {\it et al.} (2001)~\cite{ex6},
Fey (2004)~\cite{ex7},
Makii {\it et al.} (2009)~\cite{ex8},
and Plag {\it et al.} (2012)~\cite{ex9}.

\begin{figure}
\begin{center}
\resizebox{0.45\textwidth}{!}{%
  \includegraphics{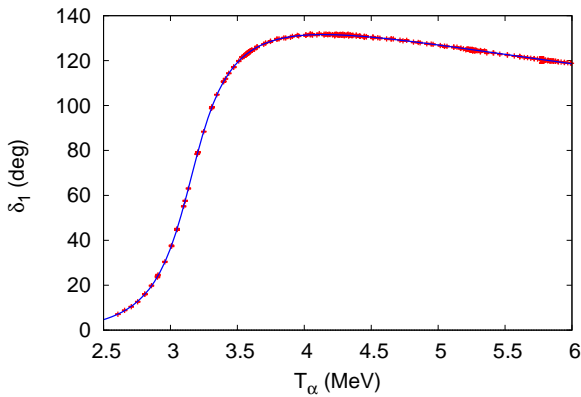}
}
\caption{
Phase shift, $\delta_1$, plotted by using the fitted effective
range parameters, $r_1$, $P_1$, $Q_1$  as a function of $T_\alpha$.
The experimental phase shift data are also displayed in the figure.
}
\label{fig;delta1}       
\end{center}
\end{figure}

As discussed in Sec.~\ref{sec:3},
we fitted the effective range parameters to the phase shift data
for $l=1$ at $T_\alpha = 2.6-6.0$~MeV and displayed the fitted
values of $r_1$, $P_1$, and $Q_1$ in Table~\ref{table:p-wave_parameters}
where the number of the data is $N=273$ and
$\chi^2/N=0.74$. The uncertainties of the fitted values
stem from those of the experimental data.
In Fig.~\ref{fig;delta1}, we plot a curve of the phase shift $\delta_1$
calculated by using the fitted effective range parameters
as a function of $T_\alpha$.
We display the experimental data in the figure as well.
One can see that the theory curve reproduces well
the experimental data at the energy range, $T_\alpha=2.6-6.0$~MeV.
Because the pole structures for the subthreshold $1_1^-$ state
and the first resonant $1_2^-$ state of $^{16}$O are well accounted
by the dressed $^{16}$O propagator, and an effect from the first 
excited $2_1^+$ state of $^{12}$C at $T_{\alpha(12)}=5.92$~MeV is appeared
weak, 
we now may regard that an irrelevant degree of freedom at high energy
for $l=1$ comes from the $1_3^-$ state of $^{16}$O 
at $T_\alpha(1_3^-) = 7.96$~MeV 
(with the width, $\Gamma(1_3^-) = 110\pm 30$~keV). 

\begin{figure}
\begin{center}
\resizebox{0.45\textwidth}{!}{%
  \includegraphics{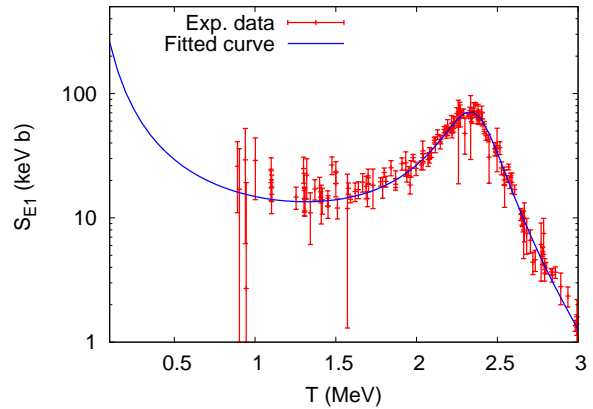}
}
\caption{
$S_{E1}$ factor plotted by using the fitted parameters
with $r_C=0.1$~fm as a function of $T$.
The experimental data are also displayed in the figure.
}
\label{fig;se1}       
\end{center}
\end{figure}

We fit the parameters, $h^{(1)R}$ and $y^{(0)}$,
to the experimental data of $S_{E1}$ at the energy range,
$T=0.9-3.0$~MeV using some values of the cutoff $r_C$
in the range, $r_C=0.01-0.35$~fm,
in the $r$ integral in $X^{(l=1)}_{(d+e)}$ in Eq.~(\ref{eq;Xde}).
The number of the data is $N=151$.
We find a significant cutoff dependence of the couplings,
$h^{(1)R}$ and $y^{(0)}$, as well as the $S_{E1}$ factor at $T_G$
when varying the short range
cutoff, $r_C=0.01 - 0.35$~fm;
as the values of $r_C$ become larger,
$\chi^2/N$ become larger while the $S_{E1}$ values at $T_G$ become smaller.
(See Table 1 in Ref.~\cite{a-18}.)
In Fig.~\ref{fig;se1}, we plot a curve of $S_{E1}$ calculated by using
the fitted parameters,
$h^{(1)R}=-0.0695(11)\times 10^4$~MeV$^4$ and 
$y^{(0)}=0.495(18)$~MeV$^{1/2}$ with $r_C=0.1$~fm 
(where $\chi^2/N=1.715$).
We display the experimental data in the figure as well.
One can see that the theory curve reproduces well
the experimental data.
In our work, we choose the results of $S_{E1}$ with
$\chi^2/N \simeq 1.7$ at $r_C\le 0.1$~fm for our estimate of $S_{E1}$
at the Gamow-peak energy, $T_G=0.3$~MeV, thus, we have
\bea
S_{E1}=59\pm 3\,\  \mbox{\rm keV$\cdot$b}\,,
\eea
where the small, about 5\%, uncertainty
stems from those of $h^{(1)R}$ and $y^{(0)}$
as well as that of the $r_C$ dependence
of $S_{E1}$ within $\chi^2/N\le 1.7$.
The previous estimates of the $S_{E1}$ factor at $T_G$ are well summarized
in Table IV in Ref.~\cite{detal-17}.
The reported vales are scattered from 1 to 340~keV$\cdot$b with various size
of the error bars.
Nonetheless it is worth pointing out that our result is about 30\% smaller
than those reported recently:
$86\pm 22$ by Tang {\it et al.} (2010)~\cite{tetal-prc10},
83.4 by Schurmann {\it et al.} (2012)~\cite{setal-12},
$100\pm 28$ by Oulebsir {\it et al} (2012)~\cite{oetal-12},
$80\pm 18$ by Xu {\it et al.} (2013)~\cite{xetal-npa13},
$98.0\pm 7.0$ by An {\it et al.} (2015)~\cite{aetal-prc15}, and
86.3 by dwBoer {\it et al.} (2017)~\cite{detal-17}.

\section{$\beta$ delayed $\alpha$ emission from $^{16}$N }
\label{sec:6}

In this section, we study the $\beta$ delayed $\alpha$ emission from $^{16}$N
by employing an EFT.
For the $R$-matrix analysis this is an important input to 
estimate the $S_{E1}$-factor at $T_G$ while for the EFT approach this is
not the case; we will discuss that though the experimental data 
of the $\beta$ delayed $\alpha$ emission are well described 
in the EFT approach, it is notably different from those in
the $R$-matrix analysis.
In the following subsections,
the formalism of $\beta$-decay and $\beta$ delayed $\alpha$ emission
from $^{16}$N is first discussed, and the decay amplitudes up to NNLO
are derived from the effective Lagrangian. After fitting parameters
to each of existing two data sets for the $\alpha$ energy distributions, 
we discuss numerical results we obtained.

\subsection{
$\beta$-decay and $\beta$ delayed $\alpha$ emission from $^{16}$N
}

$^{14}$N and $^{15}$N are stable nuclei while $^{16}$N is radioactive
whose half-lifetime is $7.13\pm 0.02$~sec
decaying through the Gamow-Teller transition 
for $\beta^-$-decay~\cite{a-pr58,bk-np64,twc-npa93}. 
It shows a first-forbidden character of the $\beta$-decay, 
and the ground state of $^{16}$N is identified as $J^\pi=2^-$. 
Branching ratios of the decaying channels to the $0_1^+$, $0_2^+$,
$3_1^-$, $2_1^+$, $1_1^-$ states of $^{16}$O and the $\alpha$-$^{12}$C breakup
channel are experimentally known as $b=0.28$, $10^{-4}$, $0.66$, $0.01$, 
$0.05$, $10^{-5}$, respectively. $p$- and $f$-waves are dominant for 
the $\alpha$-$^{12}$C breakup channel, and its $Q$ value is $Q_m'=3.257$~MeV. 
Recently, the branching ratios are updated by experiment
as $b_{\beta,11} = (5.02\pm 0.10)\times 10^{-2}$ for the bound $1_1^-$ state
and $b_{\beta\alpha}=(1.59\pm 0.06)\times 10^{-5}$ for the $\beta$ delayed
$\alpha$ emission~\cite{ketal-prl18}.  

Yields at the energy bins for the $\alpha$ kinetic energy, $T_\alpha$, for
the $\beta$ delayed $\alpha$ emission from $^{16}$N may be obtained as
\bea
n(T_\alpha) &=& CC_\eta^2 p I(T_\alpha) \left[
W_1(\eta)\left|\tilde{A}_1\right|^2
+ \frac{28}{75}W_3(\eta)\left|
\tilde{A}_3
\right|^2
\right]\,,
\nnb \\
\eea
where we have defined $n(T_\alpha)$ as a dimensionless quantity,
and $C$ is an overall constant, $C$~(MeV$^{-6}$), 
which is fitted to the experimental data later. 
In addition, a phase space integral 
$I(T_\alpha)$ is given as 
\bea
I(T_\alpha) &=& \int_0^{p_{e,max}} dp_ep_e^2 E_\nu^2 F(Z,E_e)\,,
\label{eq;I}
\eea
where $F(Z,E_e)$ is Fermi function, 
and 
\bea
E_\nu &=& Q_m'-(E_e-m_e) - T\,,
\\
p_{e,max} &=& \sqrt{(Q_m'+m_e-T)^2-m_e^2}\,,
\eea
with $T=\frac43 T_\alpha=p^2/(2\mu)$, 
and $E_e$ ($m_e$) is the electron energy (mass).
\begin{figure}
\begin{center}
\resizebox{0.27\textwidth}{!}{%
  \includegraphics{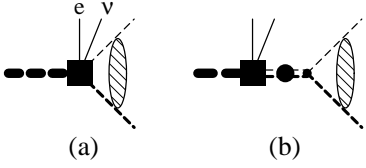}
}
\caption{
Diagrams for $\beta$ delayed $\alpha$ emission from $^{16}$N.
A thick dashed line denotes the $^{16}$N field in the initial state,
and a filled box does a weak contact vertex at which the nuclear current
and the lepton current interacts. See the captions of 
Figs.~\ref{fig;propagator} and \ref{fig;scattering_amplitude} as well. 
}
\label{fig;beta-delayed-alpha-emission}       
\end{center}
\end{figure}
The decay amplitudes, $\tilde{A}_1$ and $\tilde{A}_3$,
to the final $\alpha$-$^{12}$C state
for the $l=1$ and $l=3$ channels, respectively,
are calculated from the diagrams depicted 
in Fig.~\ref{fig;beta-delayed-alpha-emission},
and we have
\bea
\tilde{A}_1 &=& C_a^{(l=1)} + D_a^{(l=1)}\frac{p^2}{\mu^2}
+ \frac{C_b^{(l=1)} + D_b^{(l=1)}\frac{p^2}{m_O^2}}{
K_1(p) - 2\kappa H_1(p)
}\,,
\label{eq;tldA1}
\\
\tilde{A}_3 &=& C_a^{(l=3)} + \frac{C_b^{(l=3)}}{
K_3(p) - 2\kappa H_3(p)
}\,,
\label{eq;tldA3}
\eea
where we have introduced six parameters in those amplitudes;
four of them,
$C_a^{(l=1)}$, $C_b^{(l=1)}$, $C_a^{(l=3)}$, and $C_b^{(l=3)}$
are coefficients of the contact vertices for (a) non-pole and (b) pole 
diagrams for the $l=1$ and $l=3$ channels at LO. 
We also introduce two coefficients,
$D_a^{(l=1)}$ and $D_b^{(l=1)}$, for the vertex corrections for the 
diagrams (a) and (b) for the $l=1$ channel at NNLO. 
Thus, we have seven additional parameters, including the overall 
constant $C$, appearing in $n(T_\alpha)$ while the effective range parameters
appearing in the functions $K_1(p)$ in Eq.~(\ref{eq;tldA1}) 
and $K_3(p)$ in Eq.~(\ref{eq;tldA3}) 
have already been obtained in Tables \ref{table:p-wave_parameters} 
and \ref{table:f-wave_parameters}.
We use those values for the effective range parameters in the following. 

\begin{figure}
\begin{center}
\resizebox{0.13\textwidth}{!}{%
  \includegraphics{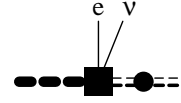}
}
\caption{
Diagram for $\beta$-decay from $^{16}$N.
See the caption of Fig.~\ref{fig;beta-delayed-alpha-emission} as well.
}
\label{fig;beta-decay}       
\end{center}
\end{figure}
In Fig.~\ref{fig;beta-decay}, a diagram for the $\beta$-decay from 
$^{16}$N is depicted.
Here we assume the perturbation expansion for the vertex correction and
include the leading ones, $C_b^{(l=1)}$ and $C_b^{(l=3)}$ only. 
Thus, the decay rates to the final $1_1^-$ and $3_1^-$ states of $^{16}$O 
are obtained as
\bea
\Gamma(1_1^-) &=& \frac85\frac{G_F^2}{(2\pi)^3}Z_1I_1C_b^{(l=1)2}\,,
\label{eq;G11}
\\
\Gamma(3_1^-) &=& \frac{28}{5}
\frac{G_F^2}{(2\pi)^3} Z_3I_1C_b^{(l=3)2}\,,
\eea 
where $I_1$ is the same integral as that in Eq.~(\ref{eq;I})
while 
\bea
E_\nu &=& m_{\rm N} - m_{\rm O}^* -E_e\,,
\\
p_{e,max} &=& \sqrt{(m_{\rm N}-m_{\rm O}^*)^2-m_e^2}\,,
\eea 
where $m_{\rm N}$ is the mass of $^{16}$N in the ground state, 
and $m_{\rm O}^*$ are the masses of the excited $1_1^-$ and $3_1^-$ states 
of $^{16}$O.  
In addition, 
$Z_1$ and $Z_3$ are wave-function normalization factors
for the $1_1^-$ and $3_1^-$ states of $^{16}$O, 
and we have
\bea
Z_1^{-1} &=& \mu\left(
r_1 + P_1 \gamma_1^2 + 6 Q_1\gamma_1^4
\right)
-4\mu\kappa \left\{\frac{}{}
H(\eta_{b1}) 
\right.
\nnb \\ &&
\left.
+ \frac{\kappa}{2\gamma_1^3}(\kappa^2-\gamma_1^2)\left[
\psi^{(1)}\left(\frac{\kappa}{\gamma_1}\right)
-\frac{\gamma_1^2}{2\kappa^2}
-\frac{\gamma_1}{\kappa}
\right]
\right\}\,,
\nnb \\ && \\
Z_3^{-1} &=& 
\mu\left(
r_3
+ P_3\gamma_3^2
+ 6 Q_3 \gamma_3^4
+ 8 R_3 \gamma_3^6
\right)
\nnb \\ && 
-4\mu\kappa\left\{
\left(
3\gamma_3^4 
- \frac{49}{18}\kappa^2\gamma_3^2
+ \frac{7}{18}\kappa^4
\right)H(\eta_{b3})
\right.
\nnb \\ && 
+ \frac{\kappa}{2\gamma_3^3}\left(
\kappa^2 -\gamma_3^2
\right)\left(
\frac14\kappa^2 -\gamma_3^2
\right)\left(
\frac19\kappa^2-\gamma_3^2
\right)
\nnb \\ && \times \left. \left[
\psi^{(1)}\left(
\frac{\kappa}{\gamma_3}
\right)
-\frac{\gamma_3^2}{2\kappa^2}
-\frac{\gamma_3}{\kappa}
\right]\right\}\,,
\eea 
where $\psi^{(n)}(z)$ are the poly-gamma function.
In the following, we fix the two parameters, $C_b^{(l=1)}$ and $C_b^{(l=3)}$, 
by using the branching ratios of the $\beta$-decay, 
and fit the remaining 
five parameters to the $\beta$ delayed $\alpha$ emission data. 

\subsection{
Numerical results for the $\beta$ delayed $\alpha$ emission 
from $^{16}$N
}

Using the experimental data for the branching ratios 
of the $\beta$-decay from $^{16}$N, 
we obtain
\bea
C_b^{(l=1)} &=& 11.4
\ \ \ \mbox{\rm MeV}\,,
\label{eq;Cb1}
\\
C_b^{(l=3)} &=& 7.13\times 10^5
\ \ \ \mbox{\rm MeV$^3$}\,.
\label{eq;Cb3}
\eea
Thus, five unfixed parameters, $C$, $C_a^{(l=1)}$, $D_a^{(l=1)}$,
$D_b^{(l=1)}$, and $C_a^{(l=3)}$ remain in $n(T_\alpha)$, 
and we fit them to the experimental data. 
Two sets of the experimental data for the $\beta$ delayed $\alpha$ emission
from $^{16}$N are
available; one is from a paper by Azuma \textit{et al.}~\cite{aetal-prc94}, 
and the other is from that by Tang \textit{et al.}~\cite{tetal-prc10}.

In Table \ref{table;parameters}, fitted values and errors 
of the parameters are displayed.\footnote{
One may check a convergence of the weak vertex correction
for the $\beta$-decay, 
the $D_b^{(l=1)}$ term, which we ignored in Eq.~(\ref{eq;G11}). 
Because of $p<Q_m$ where $Q_m$ is the $Q$ value of the $\beta$-decay 
to the ground state of $^{16}$O, $Q_m=10.419$~MeV, one has
about 1~\% correction, 
$|D_b^{(l=1)}/C_b^{(l=1)}|(Q_m/m_O)^2 =0.0109$ and 0.0099 for the two
sets of the fitted parameters. 
}
\begin{table}
\begin{center}
\begin{tabular}{c|cc} \hline
Exp. data set
 & Azuma \textit{et al.} & Tang \textit{et al.} \cr \hline
$C$ (MeV$^{-6}$) & $7.2(4)\times 10^6$ & $4.22(7)\times 10^6$  \cr
$C_a^{(l=1)}$ (MeV$^{-2}$) & $-6.9(2)\times 10^{-3}$ &
 $-9.46(5)\times 10^{-3}$ \cr
$D_a^{(l=1)}$ (MeV$^{-2}$) & 2.61(9) & 3.36(3) \cr
$D_b^{(l=1)}$ (MeV) & $-2.55(2)\times 10^5$ & $-2.297(4)\times 10^5$ \cr
$C_b^{(l=3)}$ (MeV$^3$) & $-2.65(5)\times 10^{-7}$ & $-2.46(1)\times 10^{-7}$
 \cr \hline
$\chi^2/N$ ($N$) & 4.06 (91) & 3.56 (93)  \cr \hline
\end{tabular}
\caption{
Fitted parameters. Values and errors of the parameters
are obtained by fitting to two data sets of the $\beta$ delayed 
$\alpha$ emission from $^{16}$N reported by 
Azuma \textit{et al.}~\cite{aetal-prc94} and 
Tang \textit{et al.}~\cite{tetal-prc10}. 
The numbers of the data, $N$, and values of
$\chi^2/N$ are also displayed in the table. 
}
\label{table;parameters}
\end{center}
\end{table}
We include the numbers of the data ($N$) and values of $\chi^2/N$ 
in the table as well.
\begin{figure}
\begin{center}
\resizebox{0.45\textwidth}{!}{%
  \includegraphics{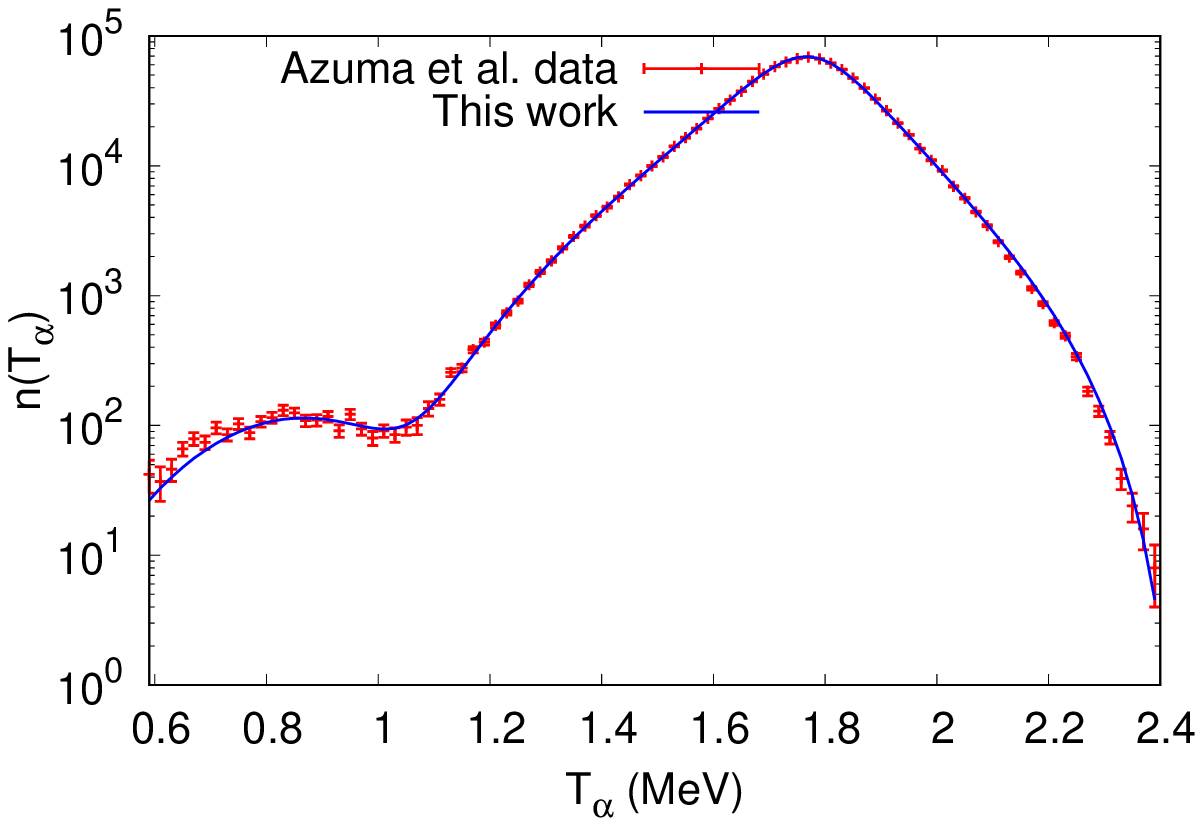}
}
\caption{
$\beta$ delayed $\alpha$ emission from $^{16}$N as a function of the 
$\alpha$ energy.
A blue curve is our fitted result, and the experimental data of 
Azuma \textit{et al.}
are included in the figure as well. 
}
\label{fig;bdad_TRIUMF}       
\end{center}
\end{figure}
\begin{figure}
\begin{center}
\resizebox{0.45\textwidth}{!}{%
  \includegraphics{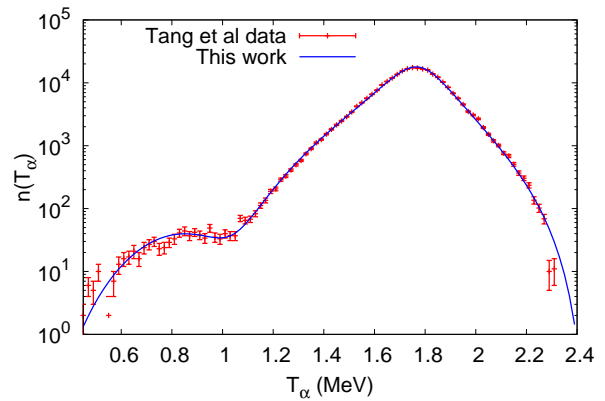}
}
\caption{
$\beta$ delayed $\alpha$ emission from $^{16}$N as a function of the
$\alpha$ energy.
A blue curve is our fitted result, and the experimental data of Tang 
\textit{et al.}
are included in the figure as well. 
}
\label{fig;bdad_Tangetal}       
\end{center}
\end{figure}
In Figs.~\ref{fig;bdad_TRIUMF} and \ref{fig;bdad_Tangetal},
curves of the $\beta$ delayed $\alpha$ emission from $^{16}$N
are plotted as a function of the $\alpha$ energy
by using the fitted parameters obtained 
in Table \ref{table;parameters}.
The experimental data are also included in the figures. 
One can see that the fitted curves reproduce the experimental data
reasonably well in the two figures though the values of the parameters
fitted from the reported data sets by Azuma \textit{et al.} and
Tang \textit{et al.} obtained in the table 
are significantly different.   
Thus, as pointed out, e.g., in Ref.~\cite{detal-17},
to see a convergence of experimental data for the $\beta$ delayed 
$\alpha$ emission from $^{16}$N is important by carrying out 
new experiments~\cite{setal-aipcp15}. 

The main peak in the figures 
appears due to the broad resonant $1_2^-$ state of $^{16}$O
while the secondary peak, as discussed before, is important to 
determine an interference pattern
between the $1^-$ levels
in the $R$-matrix or the $K$-matrix analysis. 
In the present approach, on the other hand, 
it is reproduced by an interference between the amplitudes 
from the non-pole and pole diagrams displayed
in Fig.~\ref{fig;beta-delayed-alpha-emission}.
Though one can find smaller $\chi^2 (N)$ values, 130 and 116 for 89 data points 
for background state energies,
$E_{13}=7$~MeV and 20~MeV, for $K$-matrix fit in Ref.~\cite{betal-prl93},
our $\chi^2/N$ values are comparable to those reported in a recent $R$-matrix 
analysis, $\chi^2(N)= 519 (87)$ for Azuma \textit{et al.}'s data and 
466 (88) for Tang \textit{et al.}'s data, 
i.e., $\chi^2/N = 5.97$ and 5.26, respectively, in Ref.~\cite{detal-17}.

\section{
Results and discussion
}
\label{sec:7}

In the present work, we discuss the application
of EFTs to nuclear reactions at low energies. 
We study the elastic scattering, the radiative $\alpha$ capture reaction,
and the $\beta$ delayed $\alpha$ emission from $^{16}$N
for the $\alpha$-$^{12}$C systems by constructing an EFT.
This is a typical reaction for an application of EFTs because one can
have a separation scale and many experimental data are available.
EFTs, thus, may provide us a new theoretical method, as an alternative
of $R$-matrix or potential model analysis, 
for the study of nuclear-astrophysics
where one needs to extrapolate
a reaction rate down to a low energy by fitting some parameters of theory
to experimental data measured at higher energies. 

In the study of the elastic scattering, we discuss a modification of 
the counting rules 
for an expansion around the unitary limit
due to the large suppression factor from the Coulomb
interaction at low energies; we include up to $Q^6$ order terms for 
$l=0,1,2$ and up to $Q^8$ order terms for $l=3$ in the effective range
expansion.
We find that, after fixing the first effective range term
$a_l$ by using the binding energies of $^{16}$O
and fitting the remaining effective range parameters to the phase shift
data, the broad resonances, $1_2^-$ and $3_2^-$ states of $^{16}$O 
can be described by the effective range parameters,
but the narrow resonances, $0_3^+$ and $2_2^+$ states of $^{16}$O cannot. 
As has been studied in the previous works~\cite{a-prc18,a-jkps18},
the expansion series converges well at $T_G$ while
it doesn't at the energies where the experimental data are available
(though the data are well reproduced by means of the effective range 
expansion).
In the studies for the radiative $\alpha$ capture on $^{12}$C and 
the $\beta$ delayed $\alpha$ emission from $^{16}$N, thus, we assume that the 
dressed $^{16}$O propagators for $l=1$ and $l=3$ are non-perturbative
quantities, and the capture and decay amplitudes are expanded around
them.

In the study of the radiative $\alpha$ capture reaction, 
we calculate the radiative $\alpha$ capture amplitudes 
up to NLO for the initial $p$-wave $\alpha$-$^{12}$C 
system and confirm the suppression of the $E1$ transition between the 
iso-singlet ($I=0$) nuclei. We discuss that a mixture of the $I=1$ contribution
could be introduced in the present approach by including the $p$-$^{15}$N 
and $n$-$^{15}$O channels, which are open at $T\simeq 5$~MeV and 8.5~MeV 
above the $\alpha$-$^{12}$C breakup threshold. In the present work, those states
are regarded as irrelevant degrees of freedom at high energy and integrated
out of the Lagrangian, and its effect is embedded in the contact term,
the $h^{(1)R}$ term. After fitting the two parameters, $h^{(1)R}$ and 
$y^{(0)}$ to the $S_{E1}$ data, we make an estimate of the $S_{E1}$-factor
at $T_G$ by employing an EFT.

In the study of the $\beta$ delayed $\alpha$ emission from $^{16}$N,
we calculate the $\alpha$ decay amplitudes up to NNLO.
We confirm that the primary peak of the data is accounted by
the broad resonant $1_2^-$ state of $^{16}$O while we find that 
the secondary peak is obtained from an interference between a non-pole
amplitude and a pole amplitude in the present approach.
Though the study of the $\beta$ delayed $\alpha$ emission from $^{16}$N is 
crucial in the $R$-matrix analysis, now one may see that,
except for sharing the dressed $^{16}$O propagators in the two reaction 
amplitudes, the coefficients of vertex functions for the radiative capture 
and the $\beta$ delayed $\alpha$ emission are independently
fixed by using the corresponding experimental data because 
the nuclear current for the radiative capture reaction is coupled to 
a vector current (or a minimally coupled photon) 
while that for the $\beta$ delayed $\alpha$ emission reaction
is to an axial-vector current.  
Thus, the study of the $\beta$ delayed $\alpha$ emission cannot be 
a constraint on an estimate of the $S_{E1}$-factor in our approach. 
A remarkable difference between our approach and the $R$-matrix approach
can be seen in the concept for the non-pole contribution 
for $\beta$ delayed $\alpha$ emission from $^{16}$N 
(whose diagram is displayed in the diagram (a)
in Fig.~\ref{fig;beta-delayed-alpha-emission}). 
In the present approach,
the non-pole contribution is systematically derived; 
the contact vertex functions are obtained from the effective Lagrangian 
in Eq.~(\ref{eq;Lagrangian_RC}), and the reaction amplitudes 
in Eqs.~(\ref{eq;tldA1}) and (\ref{eq;tldA3}) 
are calculated straightforwardly.
In the $R$-matrix or $K$-matrix approach, there is no non-pole contribution 
while one necessarily introduces the so-called 
``background levels"~\cite{jfhk-prc90,hfk-prc91,betal-prl93}.
Though the coefficients of the background levels are fitted to the data 
as free parameters, they play the same role in reproducing the interference 
pattern for the secondary peak. Thus, our result may 
indicate that the non-pole contributions account for an origin of 
the background levels in the $R$-matrix or $K$-matrix calculations.    
We also find that the values of the coefficients in the $\beta$ delayed
$\alpha$ emission amplitudes are quite different 
when fitting the coefficients to each of the two experimental data sets. 
Thus, to see a convergence of the experimental data
would be important by performing new experiments in the future. 

Though we have reported a first result of the radiative $\alpha$ capture on 
$^{12}$C employing an EFT, some issues remain to be explained: 
In the study of the elastic scattering, we have introduced a modification
of the counting rules from the observation of anomaly of the expansion series 
compared to the experimental data
while we have not investigated how the anomalous terms come out.
It could appear because of our assumption of the point-like particles;
a real nucleus has a finite size, and the short range contributions due to
the assumption may need to be subtracted by introducing counter terms.
In the study of the radiative $\alpha$ capture reaction, 
we reproduce the suppression
of the $E1$ transition while the non-vanishing contribution, 
the $h^{(1)R}$ term, is merely fitted to the experimental data. 
Thus it would be interesting to study
a mixture of the $I=1$ state for the $E1$ transition by including the
$p$-$^{15}$N and $n$-$^{15}$O channels in the framework of EFT. 
It is also important to include higher order terms at NNLO
in order to estimate a theoretical uncertainty of the $S_{E1}$-factor at 
$T_G$.

As we have discussed above, an application of EFTs for nuclear reactions 
would be possible, 
provided that one can choose a clear separation scale for
an observable of a reaction and the experimental data are available
to fix coefficients appearing in an effective Lagrangian. 
Thus, those nuclear reactions at low energies, which are important in 
nuclear-astrophysics, are possible candidates for the application of EFTs,
especially when the accuracy and the error estimate of a reaction are important.
It would also be interesting to study
the $E2$ transition and the cascade transitions of the radiative $\alpha$
capture on $^{12}$C at $T_G$ by employing an EFT.
A study toward this direction is now underway.

\section*{
Acknowledgements
}

The author would like to thank 
X.~D. Tang, T. Kajino, A. Hosaka, and T. Sato for useful discussions and
RCNP, Osaka University for hospitality during his stay when finalizing 
the work. 
This work was supported by
the Basic Science Research Program through the National Research
Foundation of Korea funded by the Ministry of Education of Korea
(NRF-2016R1D1A1B03930122 and NRF-2019R1F1A1040362)
and in part by
the National Research Foundation of Korea (NRF)
grant funded by the Korean government
(NRF-2016K1A3A7A09005580).

\section*{
Appendix
}
\label{appendix_a}

The elastic scattering amplitudes are calculated from 
the renormalized dressed three-point vertices and the 
renormalized dressed composite $^{16}$O propagators as
\bea
iA_l = -i\Gamma^{(l)}(k')D^{(l)}(T)\Gamma^{(l)}(k)\,,
\eea
where $k'=k$ and $T=k^2/(2\mu)$, and we have suppressed the indices from 
the Cartesian tensors.

We have the renormalized dressed three-point vertices for the 
initial and final Coulomb interaction for $l=0,1,2,3$ as
\bea
\Gamma^{(l=0)}(k) &=& y_{(0)}e^{i\sigma_0}C_\eta\,, 
\\ 
\Gamma^{(l=1)}_i(k) &=& \frac{y_{(1)}}{\mu}k_ie^{i\sigma_1}
\sqrt{1-\eta^2}C_\eta\,,
\\
\Gamma^{(l=2)}_{ij}(k) &=& \frac{y_{(2)}}{\mu^2}15e^{i\sigma_2}C_2\left(
k_ik_j -\frac13\delta_{ij}k^2
\right)\,,
\\
\Gamma^{(l=3)}_{ijk}(k) &=& \frac{y_{(3)}}{\mu^3}105e^{i\sigma_3}C_3
\nnb \\ && \times \left[
k_ik_jk_k -\frac15k^2\left(
\delta_{ij}k_k + \delta_{ik}k_j + \delta_{jk}k_i
\right)
\right]\,,
\nnb \\ 
\eea
where
\bea
C_2 &=& \frac{1}{30}C_\eta\sqrt{
(1+\eta^2)(4+\eta^2)
}\,,
\\
C_3 &=& \frac{1}{630}C_\eta\sqrt{
(1+\eta^2)(4+\eta^2)(9+\eta^2)
}\,.
\eea

We have the renormalized dressed composite $^{16}$O propagators for $l=0,1,2,3$
as
\bea
D^{(l=0)}(T) &=& \frac{2\pi }{\mu y_{(0)}^2}\frac{1}{-K_0(k)+2\kappa H_0(k)}\,,
\\
D^{(l=1)}_{i,x}(T) &=& 
P^{(l=1)}_{i,x}
\frac{6\pi\mu}{y_{(1)}^2}
\frac{1}{-K_1(k)+2\kappa H_1(k)}\,,
\\
D^{(l=2)}_{ij,xy}(T) &=& \frac32P_{ij,xy}^{(l=2)}\frac{10\pi\mu^3}{y_{(2)}^2}
\frac{1}{-K_2(k)+2\kappa H_2(k)}\,,
\\
D^{(l=3)}_{ijk,xyz}(T) &=& \frac52P^{(l=3)}_{ijk,xyz}
\frac{14\pi\mu^5}{y_{(3)}^2}
\frac{1}{-K_3(k)+2\kappa H_3(k)}\,,
\eea
where $P_{i,x}^{(l=1)}$, $P^{(l=2)}_{ij,xy}$, and $P^{(l=3)}_{ijk,xyz}$ 
are the projection operators which satisfy the relation, $P=PP$, and we have 
\bea
P^{(l=1)}_{i,x} &=& \delta_{ix}\,,
\\
P^{(l=2)}_{ij,xy} &=& \frac12\left(
\delta_{ix}\delta_{jy} 
+ \delta_{iy}\delta_{jx}
- \frac23\delta_{ij}\delta_{xy}
\right)\,,
\\
P^{(l=3)}_{ijk,xyz} &=& \frac16\left[\frac{}{}
\delta_{ix}\delta_{jy}\delta_{kz} + \mbox{\rm 5 terms}
\right. \nnb \\ && \left.
-\frac25\left(\delta_{ij}\delta_{kx}\delta_{yz} + \mbox{\rm 8 terms}
\right)
\right]\,.
\eea
In addition, the couplings, $y_{(l)}$ are redundant when one fixes
them by using the effective range parameters, conventionally one may choose
them as
\bea 
&&
y_{(0)} = \sqrt{\frac{2\pi}{\mu}}\,,
\ \ \ 
y_{(1)} = \sqrt{6\pi\mu}\,,
\nnb \\ && 
y_{(2)} = \sqrt{10\pi\mu^3}\,,
\ \ \
y_{(3)} = \sqrt{14\pi\mu^5}\,.
\eea

%

\end{document}